\documentclass[%
reprint,
onecolumn,
 amsmath,amssymb,
 aps,
aps,
floatfix,
]{revtex4-2}

\usepackage{graphicx}
\usepackage{dcolumn}
\usepackage{bm}
\usepackage{xcolor} 

\usepackage{algpseudocode}  
\usepackage{algorithm}      

\newcommand*\mean[1]{\bar{#1}}

\newcommand{\beq}{\begin{equation}}
\newcommand{\eeq}{\end{equation}}

\newcommand{\eps}{\varepsilon}

\renewcommand{\rho}{\varrho}
\renewcommand{\theta}{\vartheta}
\renewcommand{\phi}{\varphi}

\newcommand{\wegdamit}[1]{} 

\newlength{\lwveryfine}   
\newlength{\lwfine}   
\newlength{\lwnormal} 
\newlength{\lwthick}  
\newlength{\lwverythick}  

\usepackage{makecell}

\usepackage{tikz}

\newcommand{\ExternalLink}{%
    \tikz[x=1.2ex, y=1.2ex, baseline=-0.05ex]{%
        \begin{scope}[x=1ex, y=1ex]
            \clip (-0.1,-0.1) 
                --++ (-0, 1.2) 
                --++ (0.6, 0) 
                --++ (0, -0.6) 
                --++ (0.6, 0) 
                --++ (0, -1);
            \path[draw, 
                line width = 0.5, 
                rounded corners=0.5] 
                (0,0) rectangle (1,1);
        \end{scope}
        \path[draw, line width = 0.5] (0.5, 0.5) 
            -- (1, 1);
        \path[draw, line width = 0.5] (0.6, 1) 
            -- (1, 1) -- (1, 0.6);
        }
    }

\usepackage{scrextend} 
\usepackage{hyperref}   

\usepackage{pifont,mdframed}    
\definecolor{LightRed}{rgb}{1.0,.80,.79}
\definecolor{LightBlue}{RGB}{194,226,242}   
\definecolor{LightYellow}{RGB}{250,241,169}

\newenvironment{warning}{
    \par
    \begin{mdframed}[linewidth=2pt, linecolor=LightRed, backgroundcolor=LightRed]%
    \begin{list}{}{\leftmargin=1cm}{\labelwidth=\leftmargin}\item[\Large\ding{43}]}
    {\end{list}\end{mdframed}\par}

\newenvironment{hint}{
    \par
    \begin{mdframed}[linewidth=2pt, linecolor=LightYellow, backgroundcolor=LightYellow]%
    \begin{list}{}{\leftmargin=1cm}{\labelwidth=\leftmargin}\item[\Large\ding{43}]}
    {\end{list}\end{mdframed}\par}
    
\newenvironment{parameter}[2]
    {\textbf{\large{#1}} - \textit{#2} \vspace{0.2em} \begin{addmargin}[3em]{0cm}}
    {\vspace{0.7em} \end{addmargin} \ignorespacesafterend}
    
\newenvironment{function}[1]
    {\texttt{#1} \vspace{0.2em} \begin{addmargin}[3em]{0cm}}
    {\vspace{0.7em} \end{addmargin} \ignorespacesafterend}
    
\newenvironment{funcpara}[2]
    {\texttt{#1} - \textit{#2} \vspace{0.2em} \begin{addmargin}[3em]{0cm}}
    {\vspace{0.7em} \end{addmargin} \ignorespacesafterend}

\definecolor{eclipseGreen}{RGB}{63,127,95}
\definecolor{mymauve}{rgb}{0.58,0,0.82}

\usepackage{listings} 

\lstdefinestyle{myCustomMatlabStyle}{
	backgroundcolor=\color{white},   
	breakatwhitespace=false,         
  	breaklines=true,                 
 	commentstyle=\color{eclipseGreen},		
	extendedchars=true,              
	frame=single,	                   
	keepspaces=true,                 
 	keywordstyle=\color{blue},       
 	language={[Visual]Basic},                 
 	morekeywords={*,As},            
 	numbers=left,                    
 	numbersep=5pt,                   
	numberstyle=\tiny\color{gray}, 
	stepnumber=2,                    
	rulecolor=\color{black},         
	stringstyle=\color{mymauve},     
	showspaces=false,                
	showstringspaces=false,          
	showtabs=false,                  
	tabsize=2,	                   
	breaklines=true,
	morestring=[d]\"
}

\usepackage{subcaption}

\begin{document}


\title{FemtoTrack: A space-charge tracking tool for few-electron ultrashort bunches}

\author{Jan Lautenschl\"ager}
\author{Uwe Niedermayer}
 \email{niedermayer@temf.tu-darmstadt.de}

\affiliation{%
Institute for Accelerator Science and Electromagnetic Fields \\
Technical University Darmstadt \\
Schlossgartenstrasse 8, D-64289 Darmstadt, Germany
}%

\date{\today}

\begin{abstract}
Ultrafast electron experiments usually work with low-emittance few-electron pulsed beams. The structures are usually much larger than the (average) electron pulse size posing challenges to the resolution of simulations. We present the tracking code FemtoTrack, which allows tackling these multi-scale challenges in a simple manner. The computationally most heavy interpolation of the fields is treated either by a moving window or by a supplemental grid, leading to a significant speedup. Space charge is treated by direct particle-particle interaction within each bunch, where however many bunches can be simulated in the same window simultaneously. This allows to obtain statistics similar to what is obtained on the screen in an experiment. FemtoTrack is applied to two examples: An ultrafast nanotip electron source and a length scalable laser-driven electron accelerator on a microchip. In these setups, previous results have been reproduced with tremendous speedup, allowing for parameter scans similar to the tuning of an experiment.   
\end{abstract}


\maketitle

\section{Introduction}

Ultrafast electron microscopy and -diffraction (UEM/UED) has revolutionized the way we can look at macromolecules and chemical reactions therein~\cite{Zewail20104DMicroscopy}. Intense single-cycle laser pulses, obtained e.g. by chirped pulse amplification~\cite{Strickland1985CompressionPulses}, have enabled attosecond science~\cite{Krausz2009AttosecondPhysics} to discover both classical and quantum processes that show transients on unprecedented time scales. While the observations on these ultrashort time scales are mostly quantum mechanically in nature, the probing electron or laser beams are sufficiently described classically. Especially the preparation of ultrashort electron bunches, as being done in electron microscopes or similar setups, is entirely classical, with the exception of the electron emission itself. Modern electron microscopes, run in either ultrafast laser-triggered mode or CW mode, usually employ nanotip Schottky emitters, which emit electrons almost pointlike but under a large angle spread. Thus, the source emittance is almost zero~\cite{Ehberger2015HighlyTip} (close to the Heisenberg uncertainty limit); moreover the energy spread is usually less than 1~eV. The final emittance on the focus is however dominated by spherical and chromatic aberrations of the used electrostatic lenses. In a cylindrically symmetric setup, these aberrations are unavoidable due to Scherzer's theorem~\cite{Scherzer1936UberLenses,Rose2009GeometricalOptics} and can only be reduced by smart design. The simulations of such electron lens setups, in particular with short bunches of electrons that interact with each other, are a challenge inhibiting the development of simple, cost-effective UED/UEM setups on a university laboratory scale.

An important application of laser-triggered nanotip electron sources is the nascent field of research on Dielectric Laser acceleration (DLA). Transparent dielectric materials with high damage threshold allow acceleration gradients about tenfold higher than in conventional accelerators, enabling the integration of an entire particle accelerator on a microchip. The conversion of these large gradients to large energy gain requires a focusing scheme, which has recently been developed by adapting the Alternating Phase Focusing (APF) technique~\cite{Niedermayer2018Alternating-PhaseAcceleration,Niedermayer2020ThreedimensionalAccelerators}. The electrical length (length in units of the laser wavelength) of these accelerators is the same as the number of DLA cells and practically is on the order of 1000-10000. Approximating the quasi-periodicity as strict periodicity allows to treat each cell individually, which results in the semi-analytical tracking code DLAtrack6D~\cite{Niedermayer2017BeamScheme}. However, especially due to the APF phase jumps the assumption of periodic boundaries of each cell is only valid if the structures are non-resonant, and a Gibbs-phenomenon of the electric field is restricted to the phase jumps. Evaluating the consequences of this field disturbance correctly requires simulating the entire chip structure, resulting in gigabyte-sized time-harmonic electric and magnetic fields. Tracking a sub-wavelength long electron bunch in these thousands of wavelength long numerical field domains is extremely inefficient if the field data cannot be restricted to the environment of the bunch.

The challenges are thus twofold: first, the large numerically precomputed fields need to be windowed where the ultrashort electron bunch is, and second, many shots need to be considered. Usually, laser-triggered electron sources as being used in DLA experiments are run at a repetition rate of about 100~kHz. The visible result on the screen is thus only an average over a huge number of individual shots.

FemtoTrack is a dedicated electron tracker to tackle these challenges. While the program is particularly designed for low energy electrons (also in the limit $W_\mathrm{kin}\rightarrow0$), it is also relativistically correct, in order to also enable working at high energy. It is designed for ultrashort electron bunches by using either a moving window or an interpolation of the field which only requires local evaluation. The structures creating the fields may be much larger than the size of the bunches. 
To numerically efficiently determine the results of many shots, they shall be calculated all at once. Without space charge, this is given by the ergodic theorem, i.e. that the ensemble average and temporal average are equal. With space charge, the tracking in the fields can still be done for all shots simultaneously, where the particle vector is clustered. Each cluster of particles undergoes direct particle-particle space charge interaction, while they all share the same moving window.
FemtoTrack is written in Matlab~\cite{MathWorks2016Matlab} and is available as open-source for the scientific community~\cite{JanLautenschlagerFemtoTrack}. 

This paper is arranged as follows: Section~\ref{Sect:Alg} discusses the algorithm and its implementation, consisting of the two interpolators, the tracker, and the space charge model. Then Sect. \ref{Sect:App} shows the examples of an ultrafast electron source and a length scalable DLA accelerator. Finally, Sect.~\ref{Conclusion} concludes with an outlook to further applications and parallelization for high-performance computing.

\section{Algorithm}
\label{Sect:Alg}  
\begin{figure}[h]
    \centering
    \includegraphics[width=\textwidth]{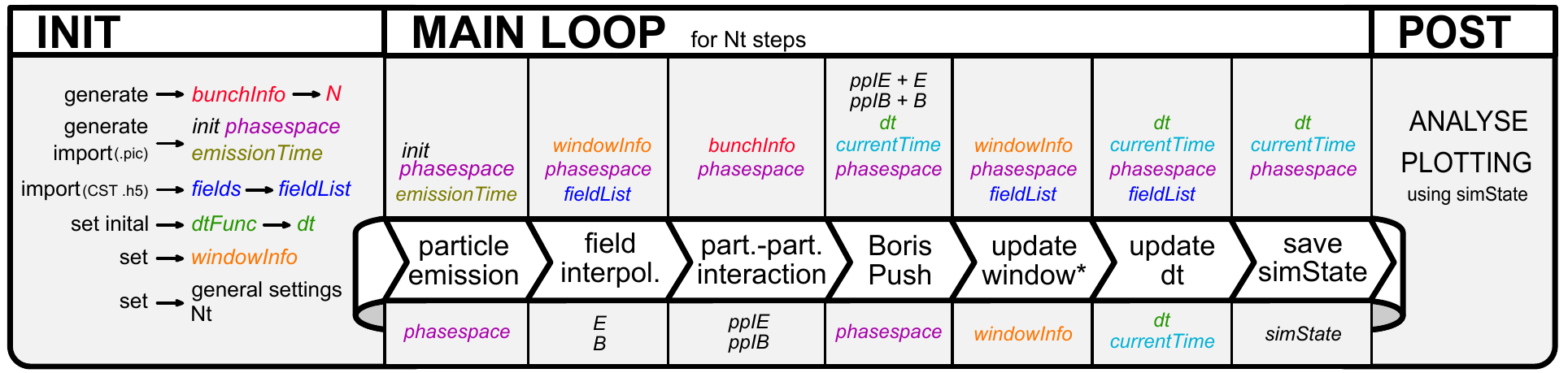}
    \caption{Schematic of the FemtoTrack algorithm. The main loop is divided into parts that are discussed individually in the text. Above each part, we note its input data, and below its output. }
    \label{fig:Algorithm}
\end{figure}

The particles are represented by one $N$-dimensional vector for each position and momentum component. This structure is called 'phasespace' and evolves with time. These structures also contain additional information describing clustering into arbitrary-sized bunches and can be copied, stored, etc. in order to represent the state of the system at an instance of time. The algorithm of FemtoTrack is schematically indicated in Fig.~\ref{fig:Algorithm}. 

Within a bunch, particles can undergo direct particle-particle interaction with one another. As long as the bunch size is small, this allows the fast, simultaneous simulation of multiple separate bunches with space charge interaction in the same external electric and magnetic fields. 

The algorithm is based on a relativistic second-order Boris method particle pusher~\cite{Boris1970RelativisticCode, Qin2013WhyGood}. To capture the particle-particle interaction in the relativistic regime, a Lorentz transformation into a local frame moving with the bunch's average velocity is performed. Consequently, the particles are almost at rest, such that the Coulomb electrostatic field suffices to compute the space charge interaction in a very good approximation. Transforming the Coulomb field back to the laboratory frame yields both an electric and magnetic field for each particle, which can be directly added to the interpolated external fields and subsequently passed to the Boris pusher.

\subsection{Particle Emission}
To realistically model particle emission due to ultra-fast laser pulses, each particle can be added to the simulation based on an individual appearance/ emission time. The resolution of the emission times is given by the simulation time step. Thus in the case that multiple particle emissions exist between two time steps, they are grouped and emitted at the next time step.

\subsection{Field interpolation}
Field interpolation is the most computationally expensive part of the tracking simulation loop. 
For each particle, the fields have to be determined at their position by interpolation from the surrounding grid nodes. As a first simplification, we limit the fields to axis-aligned, hexagonal, and three-dimensional grids, but allow non-homogeneous node spacing. For speeding up optimally in different simulation scenarios, two different interpolation techniques are implemented. 

First, a moving window is periodically updated based on the position and velocity of all particles. Using such a window limits the field data given to the interpolation function and thus gains a speed up, especially for well localized particle bunches.

\begin{figure}[h]
    \centering
    \includegraphics[width=0.5\textwidth]{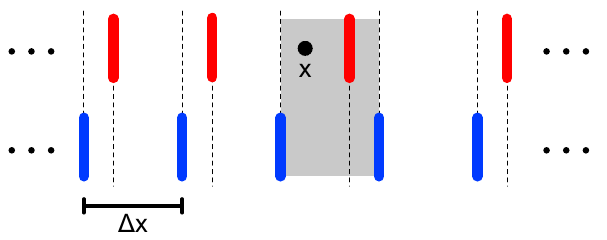}
    \caption{Sketch of the two grids, on the top the inhomogeneous grid on which the field data is defined and on the bottom the additional homogeneous grid for the faster localization of the particles. }
    \label{fig:DualGrid}
\end{figure}

The second technique is a fixed, linear field interpolator using a fast particle-to-grid node localization method. Key to this method is a lookup table consisting of $\tilde N_{i} = \frac{\max_{k}(x_{i,k})-\min_{k}(x_{i,k})}{\min_{k}(x_{i,k}-x_{i,k+1}))}$ indices for each grid direction $i \in \{x,y,z\}$, where $x_{i,k}$ are the gird node positions of the original grid with $k = 1 ... N_{i}$. The lookup table is treated as a homogeneously spaced secondary grid which is overlaid with the original grid, see Fig.~\ref{fig:DualGrid}. Each node contains the corresponding index of the node in the original grid with the next larger position value respectively. This solves the problem of finding the necessary grid nodes surrounding any point inside an inhomogeneously spaced grid by projecting it to a homogeneous grid on which the localization is negligible in computational cost, i.e. it can only be left or right from the respectively localized node on the original grid~\cite{Zhang2021FastInterpolation}. The lookup table is pre-computed for each field during the import. Note that this method reduces the interpolation effort in each time step to a small and constant amount, which can be computed in parallel for all particles. The technique is preferred over the window when the average particle spacing is much larger than the grid spacing.  

\subsection{Space Charge Model}

Since we explicitly deal with few-electron bunches here, the most efficient way of interaction is direct particle-particle interaction. As such interaction scales with $n^2$, it becomes much less efficient when the number of electrons is increased, which however is not our goal here. 

To calculate the inter-particle forces, even in the relativistic regime, we Lorentz-transform to the mean velocity~$\bm{\mean{v}}_u$ frame by applying the reversed Lorentz contraction. This is done for all particle bunches $u$ with $r=1 ... M_u$ number of particles. Since the bunches are independent, we will drop the index $u$ in the notation, keeping in mind that $M_u$ can be different for each bunch.

First, using the particle momentum $p_r$ ($r = 1 ... M$) the mean particle velocity $\bm{\mean{v}}$ and the corresponding Lorentz factor $\gamma$ is calculated 
\begin{align}
    \mean{\bm{p}} &= \frac{\sum_{r=1}^{M} \bm{p_\text{r}}}{M} \,, \\
    \gamma &= \sqrt{1 + \left(\frac{\mean{p}}{ m c_0 }\right)^2} \,, \\
    \mean{\bm{v}} &= \frac{ \mean{\bm{p}}}{m \gamma} \,.
\end{align}
Where $m$ is the particle rest-mass and $c_0$ is the speed of light. For the Lorentz-transforms the general boost matrix~\cite{Jackson1999ClassicalElectrodynamics} 
\begin{align}
    \bm{A}(\bm{v}) &= \begin{bmatrix}
        \gamma & -\gamma v_x / c & -\gamma v_y / c & -\gamma v_z / c \\
        -\gamma v_x / c & 1+\left(\gamma -1\right) v_x^2 / v^2 & \left(\gamma -1\right) v_x v_y / v^2 & \left(\gamma -1\right) v_x v_z / v^2\\
        -\gamma v_y / c & \left(\gamma -1\right) v_y v_x / v^2 & 1+\left(\gamma -1\right) v_y^2 / v^2 & \left(\gamma -1\right) v_y v_z / v^2\\
        -\gamma v_z / c & \left(\gamma -1\right) v_z v_x / v^2 & \left(\gamma -1\right) v_z v_y / v^2 & 1+\left(\gamma -1\right) v_z^2 / v^2\\
    \end{bmatrix} 
\end{align}
is determined. Second, the transformation to the comoving frame (marked with an $'$) is performed by stretching the space between particles ($x_k$ position of particles) along the mean bunch velocity using
\begin{align}
    \bm{x}'_r &= \left( \bm{x}_r - \bm{\mean{x}} \right) \cdot [A_{22}(\mean{\bm{v}}),A_{33}(\mean{\bm{v}}),A_{44}(\mean{\bm{v}})]^T ,
\end{align}
where $\mean{\bm{x}} = \frac{\sum_{r=1}^N \bm{x}_r}{M}$.

In this comoving frame, the electrostatic fields between particles are calculated
\begin{align}
    \bm{E'_{r_1,r_2}} &= \frac{e}{4 \pi \epsilon_0} \frac{\left( \bm{x_}{r_2} - \bm{x}_{r_1} \right)}{\left( x_{r_2} - x_{r_1} \right)^3} \, ,
\end{align}
with the vacuum permittivity $\epsilon_0$. To calculate the inter-particle forces in a simple way, the residual magnetic fields due to a deviation of the individual particle velocities from the mean velocity in the comoving frame are neglected. 

Third, the electromagnetic field tensor~$\bm{F}^{\mu\nu}$ with zero for the magnetic field 
\begin{align}
    \bm{F^{\mu\nu}} \left( \bm{E},\bm{B} \right) &= \begin{bmatrix}
         0  & E_x/c & E_y/c & E_z/c \\
        E_x/c &  0  & -B_z & B_y \\
        E_y/c & B_z &  0  & -B_x \\
        E_z/c & -B_y & B_x & 0 \\
    \end{bmatrix} \to
    \bm{F'}^{\mu\nu} \left( \bm{E'},\bm{0} \right) = \begin{bmatrix}
         0  & E'_x/c & E'_y/c & E'_z/c \\
        E'_x/c &  0  & 0 & 0 \\
        E'_y/c & 0 &  0  & 0 \\
        E'_z/c & 0 & 0 & 0 \\
    \end{bmatrix} 
\end{align}
is Lorentz transformed back into the lab frame using 
\begin{align}
    \bm{F} &= \bm{A}^T(\mean{\bm{v}}) {\bm{F}'\left( \bm{E'},\bm{0} \right)} \bm{A}(\mean{\bm{v}})  \,,
\end{align}
which results in an additional magnetic field due to the particle movement. 

Finally, the electric and magnetic field acting on one particle caused by all other particles is summed and superimposed with the interpolated fields at each particle's coordinates, before the total fields are fed into the Boris pusher.

\subsection{Boris Pusher}
Although we follow the original work of Boris here~\cite{Boris1970RelativisticCode}, we also exploit more modern and convenient notation~\cite{Birdsall2004PlasmaSimulation,Verboncoeur2005ParticleAdvances,Qin2013WhyGood}. 
The algorithm is the explicit, second-order, relativistic Boris pusher, which advances the particle position $x$ and momentum $p$ in the given electric and magnetic fields $E$ and $B$ evaluated at the particle position by the interpolator. The index $j$ denotes the discretized time and individual particles have charge $q$ and mass $m$. The particles move in one time step $\Delta t$ according to the equations as to be evaluated in respective order
\begin{equation}
    \label{eq:Docu_Boris_1}
    \begin{aligned}
      \bm{p}^-_j = \bm{p}_{j-1} + \frac{q \bm{E}_j \Delta t_j}{2} \, , \\
     \gamma_j = \sqrt{1 + \left(\frac{p^-_j}{ m c_0}\right)^2} \, , \\
      \bm{t_j} = \frac{q \Delta t_j \bm{B_j}}{2 m \gamma_j}  \, , \\
     \bm{s}_j = \frac{2 \bm{t}_j}{1 + t_j^2} \, , \\
      \bm{d}_j = \bm{p}^-_j + \left( \bm{p}^-_j \times \bm{t}_j \right),   \, \\ 
     \bm{p}^+_j = \bm{p}^-_j + \left( \bm{d}_j \times \bm{s}_j \right) \, , \\
          \bm{p}_j = \bm{p}^+_j + \frac{q \bm{E}_j \Delta t_j}{2}  \,   
    \end{aligned}
\end{equation}
where $\bm{t}$, $\bm{s}$ and~$\bm{d}$ are intermediate step vectors, abbreviating the state between two half electric field pushes and non-bold symbols indicate magnitudes. The electric field $E$ acts on the momentum of the particles for only half a time step, becoming~$p^-$. Then the magnetic field applies a rotation with the full time step, and finally, the second half electric field kick is applied. With the total kick in the time step $\Delta t_j$, the total push becomes 
\begin{equation}
    \label{eq:Boris_2}
    \bm{x}_j = \bm{x}_{j-1} + \Delta t_j \cdot \frac{\bm{p}_j}{\gamma_j m}
\end{equation}
which completes the explicit scheme.

\subsection{Window and Time Step Adaption}
The window is updated according to the particle distribution to minimize the active field grid cells used for interpolation. The time step can be chosen arbitrarily, and also arbitrarily adapted during the simulation. A convenient way to choose the time step is by following the CFL condition~\cite{R.CourantK.Friedrichs1928OnPhysics}, which can be written as 
\begin{align}
    \Delta t = k_\mathrm{CFL} \frac{\min_\text{i,k} ( \Delta x_\text{i,k} ) }{ v_\text{max}}
\end{align}
where $k_\mathrm{CFL}\leq 1$ is the adjustable CFL number, $\Delta x_\text{i,k}$ the cell size along $i = \{x,y,z\}$ and $v_\text{max}$ the fastest particle velocity. Note that the minimum is taken only over the volume covered by the particle distribution, which avoids spoiling the time step by irrelevant small grid edges.

\section{Applications}
\label{Sect:App}

FemtoTrack is applied to two test cases; one example is an ultrafast, low emittance electron source as detailed in~\cite{Leedle2022HighSources} and another example is a nano-photonic DLA accelerator chip as in~\cite{Niedermayer2021DesignChip}. Both scenarios have in common, that extensive multi-dimensional parameter studies need to be performed, in order to find the optimal initial conditions for high beam brightness. The difference between the two is that only the source requires consideration of space charge, since only at both high charge density and low energy this is a concern. In the following, we report extensive fast parameter studies done with FemtoTrack, that were previously impossible to perform due to the insufficient speed of the used generic commercial software. 

\subsection{Ultrafast electron source}
The electron gun called "Glassbox" is used for a variety of DLA experiments at Stanford University~\cite{Leedle2022HighSources}. 

\begin{figure}[h]
    \centering   
    \includegraphics[width=0.7\textwidth]{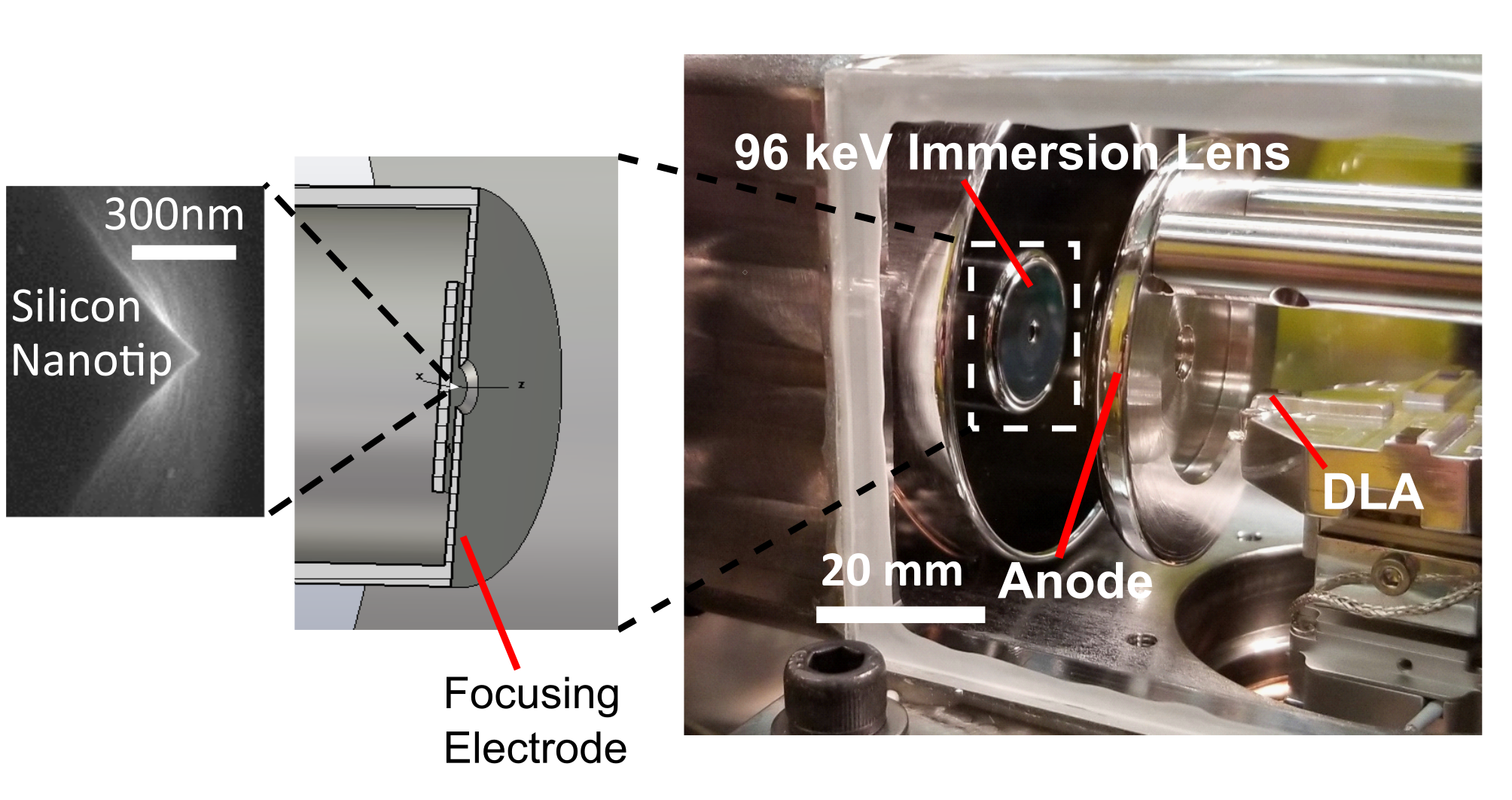}
    \caption{Stanford Glassbox setup. Picture adopted from~\cite{Niedermayer2022BeamAccelerators}.}
    \label{fig:Glassbox}
\end{figure}

\begin{figure}[ht!]
    \centering
    \includegraphics[width=0.45\textwidth]{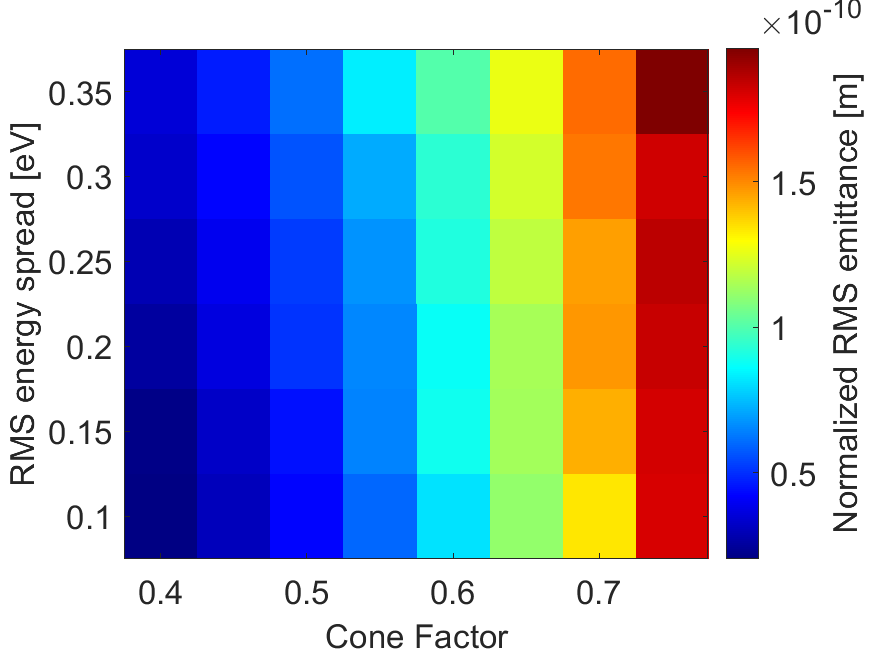}
    \includegraphics[width=0.45\textwidth]{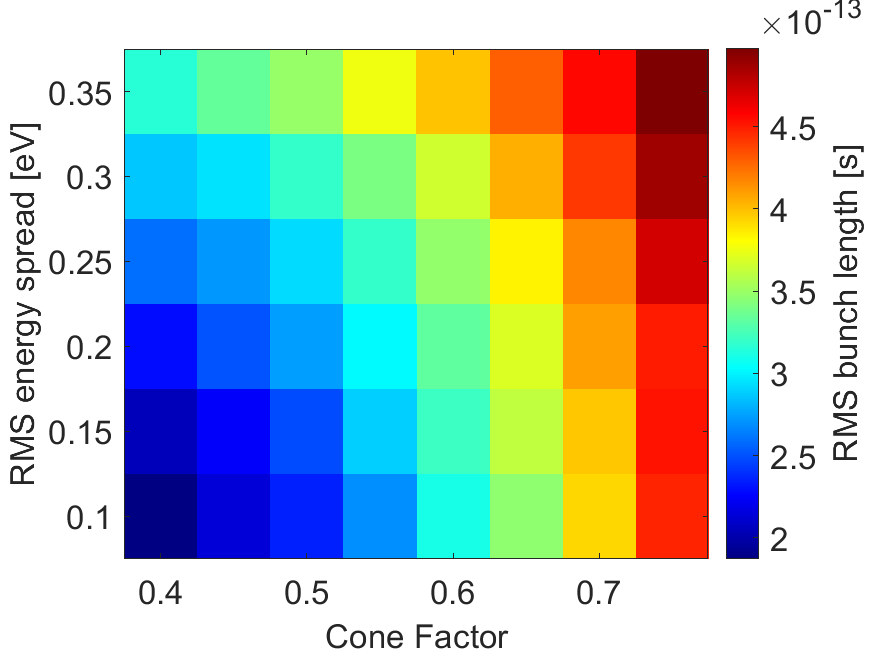}
    \caption{Source parameter sweep for the Stanford Glassbox.}
    \label{fig:GlassboxSweep}
\end{figure}

\begin{figure}[b]
    \centering
    \includegraphics[width=\textwidth]{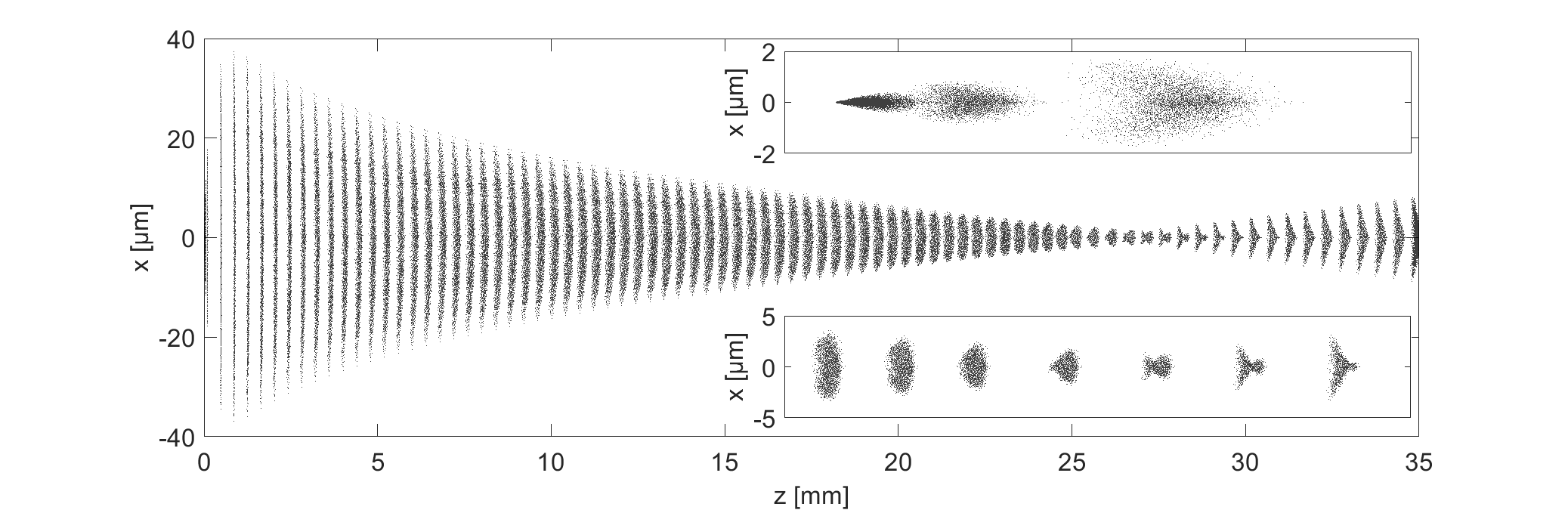}
    \includegraphics[width=\textwidth]{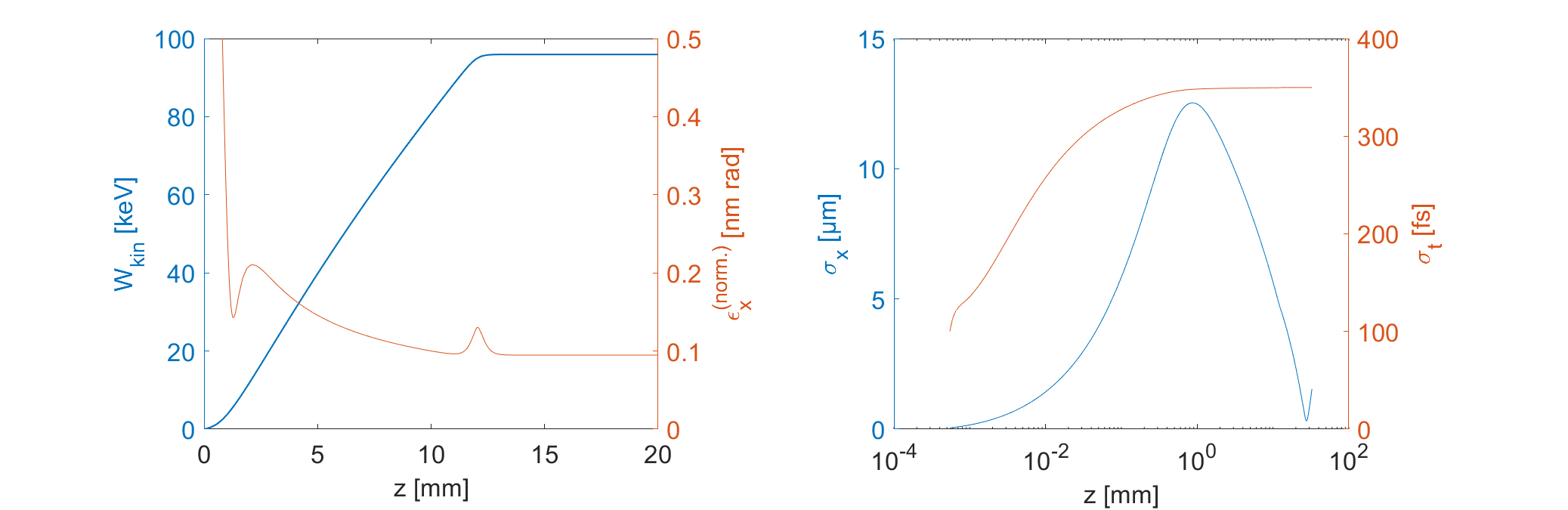}
    \caption{Results of a single shot with initial parameters fitting the bunch length and emittance measurement at the focus and at negligible total current.}
    \label{fig:GlassboxZeroCurrent}
\end{figure}

As depicted in Fig.~\ref{fig:Glassbox}, it consists of a silicon nanotip emitter, followed by a focusing electrode (sometimes called immersion lens~\cite{Hirano2020AAccelerator}) made of Silicon-Carbide and finally a grounded polished steel anode. The emission takes place at -96~kV from a silicon nanotip~\cite{AndrewCeballos2019SiliconAccelerators}, where the immersion lens electrode is another 1.847~kV lower in potential, to repel the electrons towards the optical axis. The emitter itself is modeled as a hemisphere of 20~nm radius, on an elevated cone of 400~nm height. The properties of the initial electron distribution are widely unknown, and can only be fitted to experimental observations, which describe the bunch train at the focus located at about 3~cm distance from the cathode. In particular, the emission cone angle and the initial energy spread are crucial parameters, as they directly probe the spherical and chromatic aberrations of the lensing system. Consequently, they dominate the emittance and the bunch length at the focus, respectively. However, the opposite effects, i.e. the effect of the energy spread on emittance and the effect of the emission cone angle on the bunch length are smaller but still non-negligible.
The initial distribution can be modeled arbitrarily in FemtoTrack. Here we chose a cosine distribution in the angle $\theta$ to the axis of the hemisphere and the emission takes place perpendicularly to the hemispherical emitter surface. The angular width of the distribution is scaled by the 'Cone Factor' ranging from 0 to 1, where 1 indicates the bottom width of the cosine distribution to be equal to the entire emitter hemisphere, i.e. the distribution starts at the equator. The energy distribution is chosen as approximately Gaussian with a mean value of 1.5~eV, where negative sample values are dropped and the distribution is re-normalized. 
Figure~\ref{fig:GlassboxSweep} shows a two-dimensional sweep over cone factor and energy spread, which allows matching emittance and bunch length in the focus to the measured values of $\eps_n=97$~pm and $\sigma_t=352$~fs. These values can be approximately matched by the choice of cone factor 0.6 and energy spread 0.25~eV (RMS). The complete simulation results with these parameters are shown in Fig.~\ref{fig:GlassboxZeroCurrent}, where the emittance and bunch length evolution along the beam line is plotted. Note that for better visualization the z-axis was chosen logarithmic at the beam size and bunch length plot.

\begin{figure}[h]
    \centering
    \includegraphics[width=0.5\textwidth]{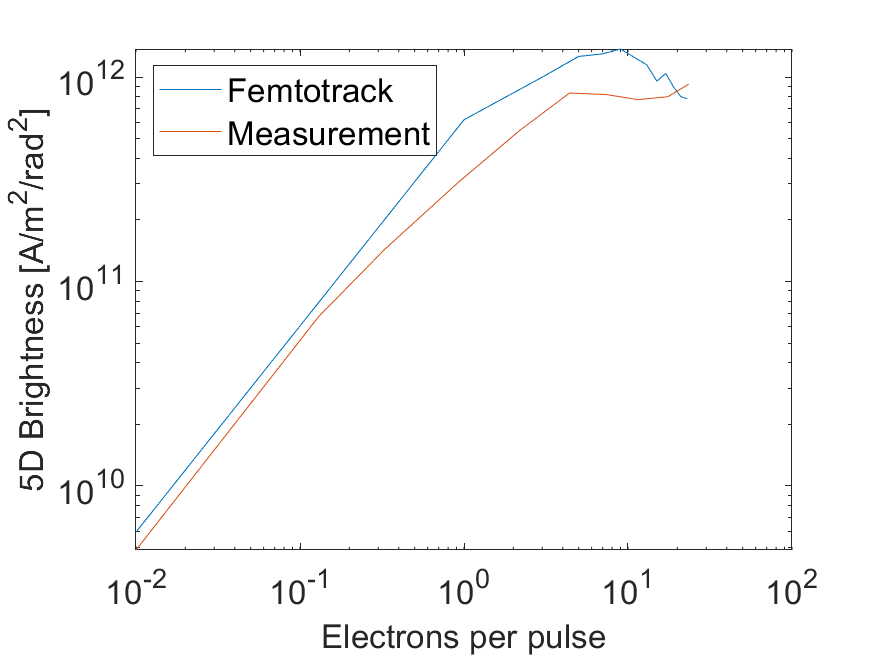}
    \caption{Brightness over bunch charge for the Stanford Glassbox ultrafast electron source. Measurement curve from~\cite{Leedle2022HighSources}.}
    \label{fig:GlassboxBrightness}
\end{figure}

Switching on the space charge module, we can simulate the emittance and brightness increase for a higher number of electrons per laser pulse. Figure~\ref{fig:GlassboxBrightness} summarized this in the brightness curve, where the brightness is defined as $B=Ne/4\pi^2\eps_n^2\sigma_t$. The simulation result is compared to the measurement as taken from~\cite{Leedle2022HighSources}.

As visible, there is an acceptable agreement. Sources of deviations are not only the measurement uncertainty, but also the errors in the approximation of the initial distribution function, which is subject to educated guess assumptions and can surely be doubted. To gain more confidence in both the measurement and simulation results, more measurement data on different nanotips, as well as more precise knowledge of their geometry would be required, which could then again be simulated in FemtoTrack. 

\subsection{3D APF DLA}
The 3D APF DLA accelerator example uses the frequency domain field import of FemtoTrack. 
\begin{figure}[h]
    \centering 
    \includegraphics[width=0.55\textwidth]{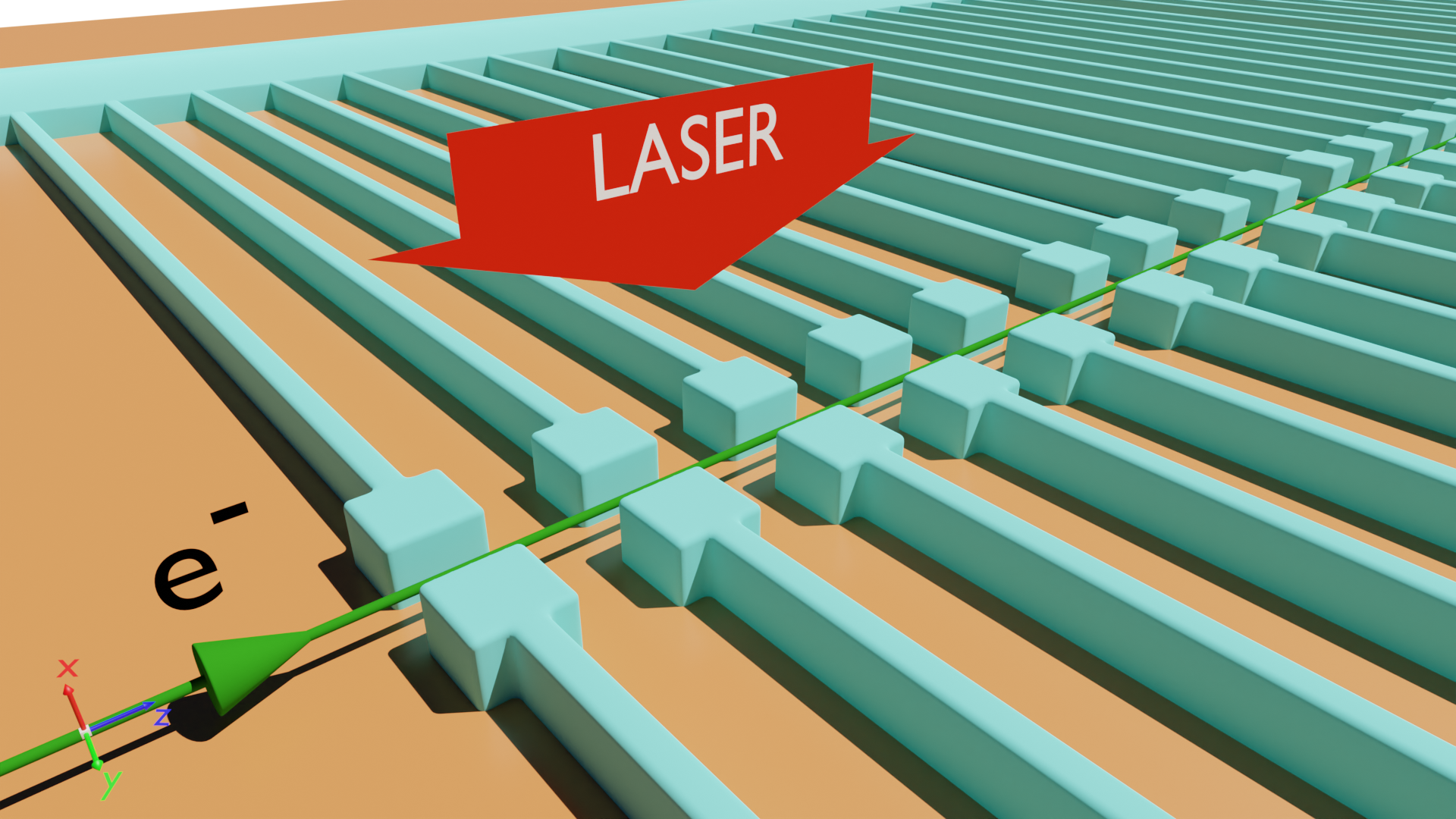}
    \caption{3D APF DLA on SOI structure, designed as a scalable accelerator from 26.5~keV to 53~keV. The electron beam channel is a rectangle of only 200~nm by 220~nm in size. Upon leaving the channel, electrons are considered as lost. Picture adopted from~\cite{Niedermayer2021DesignChip}.}
    \label{fig:SOIDLAl}
\end{figure}
\begin{figure}[h]
    \centering
    \includegraphics[width=0.47\textwidth]{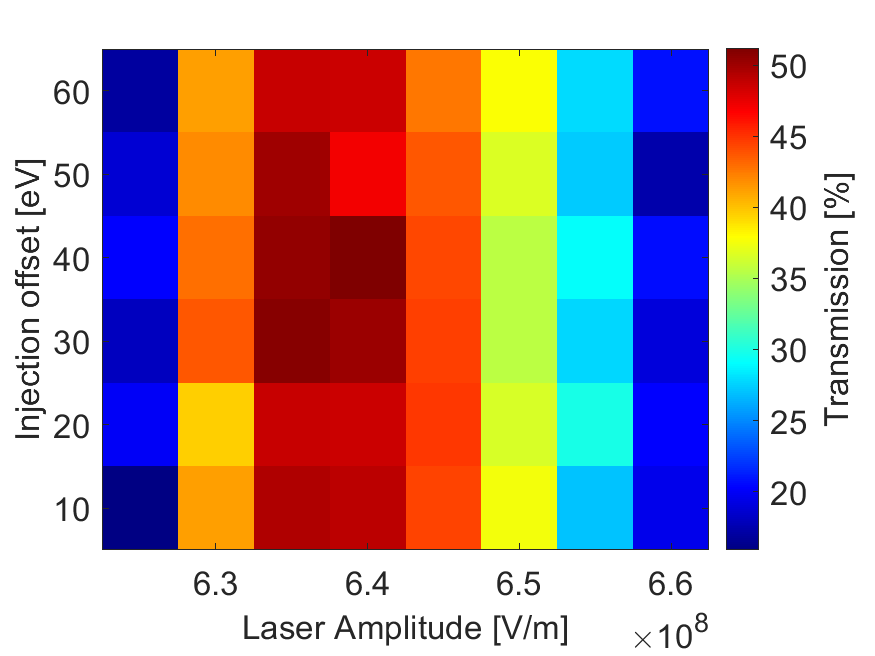}
\includegraphics[width=0.47\textwidth]{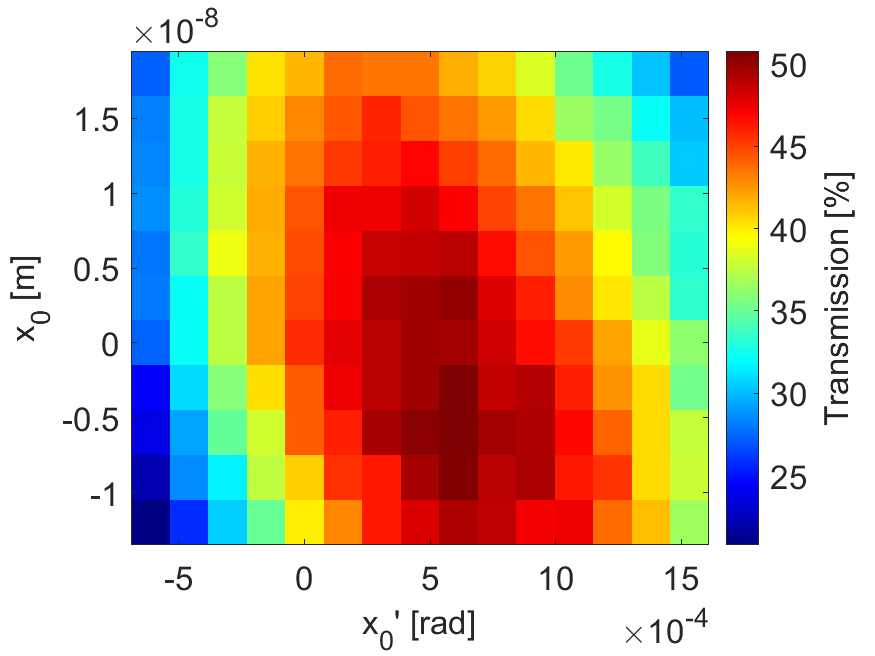}
\caption{Laser amplitude and injection energy offset scan (left) and injection position and angle scan (right).}
    \label{fig:2Dscans}
\end{figure}
The structure follows the last model presented in~\cite{Niedermayer2021DesignChip}, which is designed consisting of 220~nm high rectangular pillars on an SOI wafer, see Fig.~\ref{fig:SOIDLAl}. While the geometry of the structure is just copied here, i.e. the frequency domain field simulation done by CST Studio~\cite{Dassault-Systems2021CSTSuite} is identical to~\cite{Niedermayer2021DesignChip}, the initial conditions of the beam can be optimized by sweeping over many variables. Note that by convention the vertical coordinate is $x$, in which the structure is not symmetric, while the horizontal coordinate $y$ is mirror-symmetric about the $y=0$ plane. However, ideally, the structure design should provide as good as possible symmetry in the vertical electromagnetic fields as well. Deviation from this symmetry leads to beam loss and needs to be carefully assessed by simulations. Therefore, in this example, we take the transmission (electrons in / electrons out) as the quantity of interest. 
First, we look at bunched beams, as they are somewhat ideal for the APF structure design. The bunch parameters are: Gaussian 6D profile with geometric RMS emittance 10~pm (beam size and divergence matched to the initial beta function of the accelerator chip) and RMS bunch length 10~nm, with matched energy spread according to the initial longitudinal beta function. Under ideal circumstances, these parameters should provide transmission in excess of 50\%.  
The 5 variables initial beam position $x_0$, beam angle $x_0'$, injection energy offset, the laser amplitude, and phase are being scanned by FemtoTrack. The injection energy offset also corrects the error due to the injection window, i.e. the particles suddenly appear when the fields are already there, thus receiving an artificial energy kick in violation of the Lawson-Woodward theorem. 
\begin{figure}[h]
    \centering
    \includegraphics[width=0.47\textwidth]{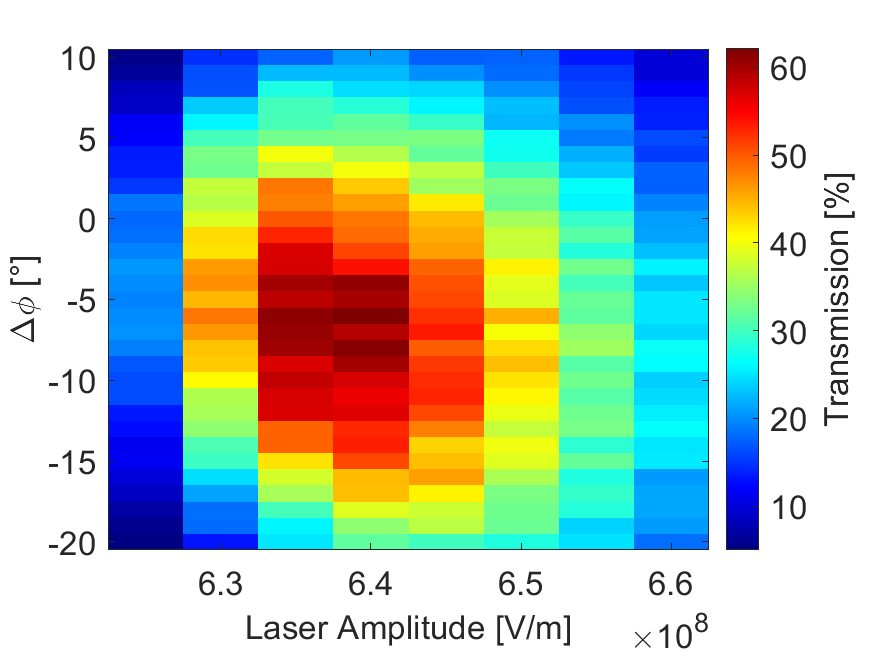}
    \includegraphics[width=0.47\textwidth]{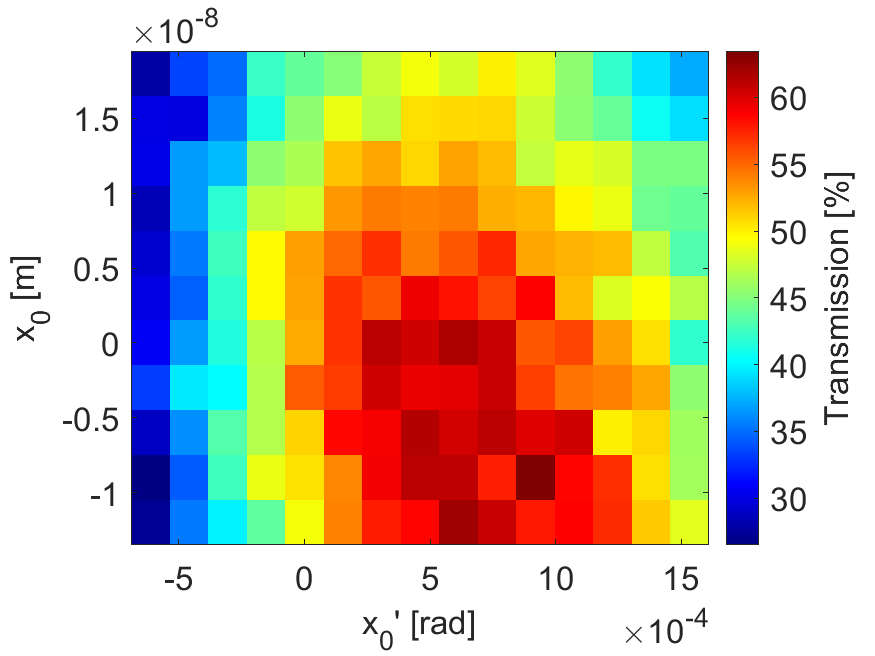}
    \caption{2D Scans of bunched beams over Laser amplitude and phase (left) and once again over the vertical injection with the optimal 637.5~MeV/m laser amplitude and $-6^\circ$ phase adjustment.}
    \label{fig:2Dscans_phasecorrected}
\end{figure}
\begin{figure}[h]
    \centering
    \includegraphics[width=0.47\textwidth]{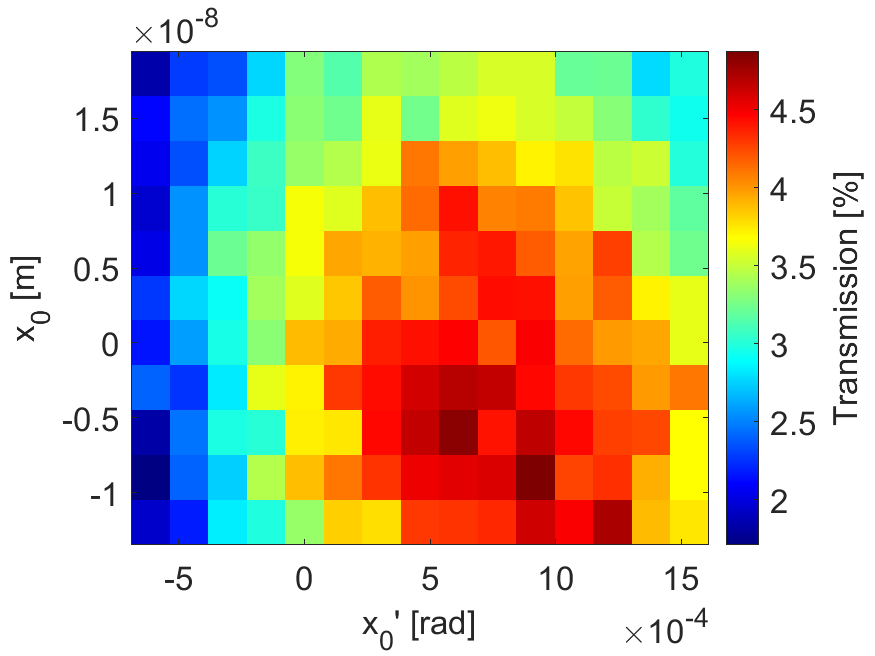}
    \caption{Scan of the vertical injection coordinates with an unbunched beam, similar as in recent experiments.}
    \label{fig:2Dscan_unbunched}
\end{figure}
Figure~\ref{fig:2Dscans} shows the transmission for a scan at the nominal injection phase, indicating optimal parameters at about 637.5~MV/m laser strength and 30~eV injection offset. Moreover, the optimal injection position is at about $x_0=5$~nm with an angle of about $x_0'=0.5$~mrad. 
Scanning over the phase reveals, that the highest transmission occurs when slightly deviating from the design injection phase by $-6^\circ$, see Fig.~\ref{fig:2Dscans_phasecorrected} (left). 
With the correct phase, and the correct injection coordinates, transmission above 60\% can be achieved (Fig.~\ref{fig:2Dscans_phasecorrected}, right). 
Finally, when turning towards a realistic setup of an experiment without attosecond bunching, i.e. an experiment with an unbunched beam, obviously the phase dependence disappears, and transmission of almost 5~\% occurs at optimal coordinates, see Fig.~\ref{fig:2Dscan_unbunched}. A similar scan is performed in the alignment of an experiment, thus one can directly compare the results and understand the parametric dependencies.

\section{Conclusion}
\label{Conclusion}
Following up on experiments requires pragmatic approaches to simulations. The tracking code FemtoTrack is such a simple and pragmatic solution to ultrafast electron dynamics. While the code is tested here only for subrelativisitc examples, it is also relativistically consistent and can treat electron bunches of arbitrary energy with and without space charge. The simplicity of the space charge solver however restricts the approach to few-electron bunches, where the statistics are obtained from many such bunches being simulated in a commonly localized part of the much larger external fields. Increasing the number of electrons is problematic here, since the computational load scales with $n^2$; more advanced techniques as e.g. the Fast Multilevel Multipole  (FMM) method~\cite{Schmid2021GitterfreieTeilchenstrahlen} can be applied here. Such methods can in the best circumstances reduce the computational load to the order of $n$, allowing the simulation of bunches with an extremely high charge at low energy as e.g. in RF photoinjectors. The low bunch intensities, as being used in UEM/UED can be treated very well with FemtoTrack as it was demonstrated in the two examples of a nanotip-emitter gun and an on-chip DLA scalable acceleration structure. Due to its simplicity and fast computation time, FemtoTrack simulations can mimic the tuning of an experiment by extensive parameter scans. This serves to better understand why particular parameter choices in an experiment work well and help to make experiments more reproducible. The FemtoTrack Code is available here~\cite{JanLautenschlagerFemtoTrack} and a documentation is found in the Appendix \ref{chp:Docu}.

\newpage
\section{Acknowledgements}
This work is funded by the German BMBF (Grant No. 05K19RDE ), LOEWE Exploration by the state of Hesse, and the Gordon and Betty Moore Foundation under Grant No. GBMF4744 (ACHIP).

\bibliography{./references}
\bibliographystyle{unsrtnat}


\newpage
\appendix
\section{Documentation} \label{chp:Docu}

\begin{figure}[h!]
    \centering
    \includegraphics[width=15cm]{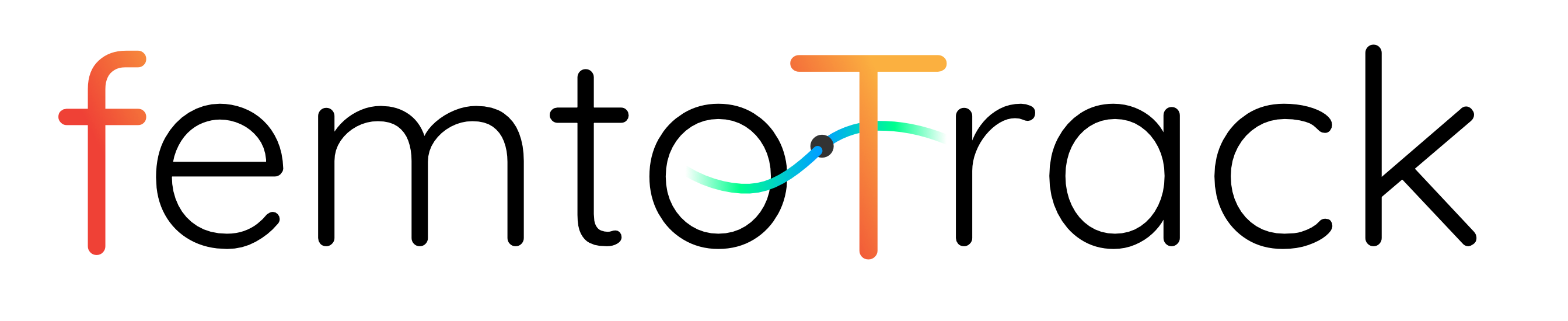}
    \LARGE \textbf{Documentation for Version 4}\normalsize \\~\\ \today
\end{figure}

\vspace*{2mm}

\begin{figure}[h!]
    \centering
    \href{https://git.rwth-aachen.de/jan.lautenschlaeger/femtotrack}{Git Lab femtoTrack: https://git.rwth-aachen.de/jan.lautenschlaeger/femtotrack \ExternalLink}
\end{figure}

\thispagestyle{empty}

\vspace*{2mm}

\newpage
\subsection{Explanation of terms} \label{chp:expTerms}

\begin{table}[h!]
    \centering
    \caption{Short explanation of commonly used terms and definitions}
    \begin{tabular}{r|l}
        \hline
        \thead{Term} & \thead{Description} \\ 
        \hline \hline
        (alive) particle & \makecell{single tracked particle in the simulation, e.g. an electron} \\  
        dead particle & \makecell{particle that got removed from the simulation} \\ 
        bunch & \makecell{a collection of particles which can interact with one another} \\  
        phase space & \makecell{the collected information about the position and momentum of all particles} \\  
        FTLib & \makecell{the folder containing all relevant files for exp. the solver and initial phase space generators} \\ 
        simState & \makecell{a data structure containing all the information generated by the simulation, \\for additional information see Chap. \ref{chp:Docu_SimStat}} \\  
        solver & \makecell{the complete core simulation code, containing among other things the field interpolation, \\particle interaction, and Boris kick, for more info see Chap. \ref{chp:Docu_Solvers}} \\
        (simulation) step & \makecell{one loop of the solver, advances the simulated time by the time step \texttt{dt}} \\
        simulation time & \makecell{real time the solver needs for the computation} \\
        simulated time & \makecell{the time computed for the particles} \\
        mesh/grid & \makecell{the three dimensional, hexagonal grids, fields are defined on} \\ 
        mesh node & \makecell{one mesh intersection in the mesh/grid} \\ 
        window & \makecell{a collection of indices marking a sub-volume of a field based on the underlying mesh} \\
        emission time & \makecell{point in time greater than zero a particle is added to the simulation} \\
        box & \makecell{matrix of $(3,2)$ floats describing a axes aligned rectangle \\with $(min X, max X; min Y, max Y; min Z, max Z)$} \\
        \hline
        float & a decimal number \\  
        int & a integer \\  
        String & \makecell{a collection of characters, in Matlab a char vector marked by ' \\(using a String in Matlab - marked by " - may work, but can lead to problems)} \\ 
        list & \makecell{in Matlab called a cell array, marked by curly brackets \{\}} \\
        vector/matrix & \makecell{in Matlab marked by square brackets []} \\
        struct & \makecell{grouping related data using containers called \textit{fields}; access data via: structName.fieldName} \\
        \hline
    \end{tabular}
\end{table} 
\begin{hint}
    The Matlab code is tested with Matlab R2021b. 
\end{hint}

\begin{hint}
    The current version of the code is tailored for electrons, but could easily be adapted to arbitrarily charged particles. 
\end{hint}

\subsubsection{Overview} \label{chp:overview}

An Example of the part the user integrates with is contained in the Matlab file \url{FemtoTrack_main_v4.m}. This file primarily consists of the user-defined initial simulation settings, the call to the solver, and the plotting functions. The user is advised to copy the necessary code into a separate, problem-specific main file.

The simulation framework is structured so that the user defines the bunches, the initial phase space, and the acting electrical and magnetic fields in addition to the general settings for the solver. For example, for the initial phase space, particles can be randomly distributed in a sub-region and fields are imported from one or multiple HDF5 files. The phase space and list of fields are then passed to one of the implemented solvers, all based on the \href{https://www.particleincell.com/2011/vxb-rotation/}{Boris method \ExternalLink}. The solver simulates all bunches simultaneously, but any particle-particle interaction are only between particles in the same bunch. After completing the simulation, the solver returns a data structure (called \texttt{simState}) containing all generated data, which can then be used for a more detailed analysis or visual representation of the simulation. 

It follows an overview of all setting options; for more details about the solver and other functions contained in the FTLib see Chap. \ref{chp:Docu_InsideFTlib}.

\newpage
\subsection{Quick Start Guide (with CST)} \label{chp:QuickStart}

This section briefly describes how a simulation could work from start to finish. In the following, it is assumed, that the simulation software CST STUDIO SUITE by Dassault Systemes is used to create the electric and magnetic field data. 

\begin{enumerate}
    \item Create the model in CST and add an HDF5 field export as post-processing steps. Simulate the field.
    \item Locate the HDF5 files.
    \item It is recommended to copy the necessary femtoTrack files, the \texttt{FTLib} folder, and the main file, to a separate working folder. 
    \item Open the main Matlab file.
    \item Import the exported HDF5 field and set the initial particle distribution. 
    \item Set the time step \texttt{dt} and the number of time steps \texttt{Nt}.
    \item Check all other settings - plotting, solver, etc. The default values should give a good initial result, which can be adjusted for the given problem in subsequent runs.
    \item Start the script. Information about the created particles and imported fields should now appear in the command line sub-window.
    \item Wait for the simulation to end. The given estimated duration of the simulation is only very rough and depends heavily on the problem.
    \item Finally, the selected plots appear. Tip: It is not necessary to run the simulation again if only a new plot needs to be created, since the simulation data (\texttt{simStat}) is retained in the workspace after the script has finished.
\end{enumerate}

\subsection{Main Script Settings} \label{chp:Docu_Settings}

\begin{hint}
    In the following sections the notation \{x1,x2,x3,...\} is to be understood that each element exists individually and when applicable, [y1,y2,y3,...] represent a list of all possible options, where a plus (+) between lists notes the existence of all combinations between the lists. Example: [a,b,c]+[1,2] $\to$ [a1,b1,c1,a2,b2,c2]. Caution: Do not confuse with example code sections where the normal Matlab notation applies.
\end{hint}

\begin{hint}
    All physical values are given in SI units. The only exception is electronvolt (\texttt{eV}) used for energy, instead of Joules (\texttt{J}). If this is the case, it is marked.
    
    \begin{tabular}{c|c|c} 
        \hline
        quantity & unit & symbol \\ 
        \hline \hline
        time & second & s, sec \\  
        length & meter & m \\  
        mass & kilogram & kg  \\  
        electric current & ampere & A \\ 
        energy & joule & J \\ 
        energy & electronvolt & eV \\ 
        \hline
    \end{tabular}
\end{hint}

\vspace*{5mm}

\begin{parameter}{bunchInfo}{struct}
    A struct containing all information about the grouping of particles.

    \vspace{0.5em}
    \begin{parameter}{phasespace.N}{int}
    The total number of particles has to match the number of particles in the phase space.
    \end{parameter}
    \begin{parameter}{phase space.bunchSizes}{$(M,1)$ int}
    Number of particles in the corresponding bunch, also sometimes called $NB$.
    \end{parameter}
    \begin{parameter}{phasespace.grouping}{$(M,2)$ int}
    The start and end indices of each bunch in the phase space vectors.
    \end{parameter}
    \begin{parameter}{phasespace.numBunches}{int}
    The number of bunches, also sometimes called $M$.
    \end{parameter}
    
    \vspace{0.2em}
    \begin{function}{bunchInfo = genEqualBunches(NB, M)}
    Generates the information for $M$ equally sized bunches with $NB$ particles each.
    
    \vspace{0.2em}
    \begin{funcpara}{NB}{int}
    The number of particles per bunch.
    \end{funcpara}
    \begin{funcpara}{M}{int}
    The number of bunches.
    \end{funcpara}
    \textit{EX}: \texttt{bI = genEqualBunches(50, 10)}
    \end{function}
\end{parameter}

\begin{parameter}{phasespace}{struct}
    The initial information about the positions, momenta, and emission times of all $N$ particles. It is represented as a struct with the fields \url{.N}, \url{.pos}, \url{.mom} and \url{.emissionTime}.

    During the initial creation, only \url{.N}, \url{.pos}, \url{.mom} and \url{.emissionTime} have to be set by the user, all other fields of the struct are generated by the solver during the simulation and thus appear in the \texttt{simState}. 
    
    \vspace{0.5em}
    \begin{parameter}{phasespace.N}{int}
    The total number of particles has to match the number of particles in the bunchInfo.
    \end{parameter}
    \begin{parameter}{phasespace.emissionTime}{$(N,1)$ float}
    Per particle time the particle is added to the simulation with the given position and momentum. The time has to be greater than zero.
    \end{parameter}
    \begin{parameter}{phasespace.mom}{$(N,3)$ float}
    The relativistic momentum of the particles in x,y, and z.
    \end{parameter}
    \begin{parameter}{phasespace.pos}{$(N,3)$ float}
    The position of the particles in x,y, and z.
    \end{parameter}
    \begin{parameter}{phasespace.alive}{$(N,1)$ bool}
    Describes the state of the particle, if it is true and the particle was emitted, then it is actively simulated. If it is false, that particle has died and is removed from the simulation.
    \end{parameter}
    \begin{parameter}{phasespace.wasEmitted}{$(N,1)$ bool}
    Describes the state of the particle, if true the particle has been added to the simulation.
    \end{parameter}
    \begin{parameter}{phasespace.killReason}{$(N,1)$ int8}
    Notes the reason why a particle died or was removed from the simulation. \\
    The currently supported options are: \\
    $-1 \to$ particle left all fields, \\
    $-2 \to$ particle too slow, \\
    $-3 \to$ removed by killBox.
    \end{parameter}
    
    \vspace{0.2em}
    \begin{function}{phasespace = genRandInitPhasespace(N, initPosRange, initMomRange, initEmissionTimeRange)}
    Generates an initial, random distribution of \texttt{N} particles in a given position and momentum regime with a random spread of emission times.
    
    \vspace{0.2em}
    \begin{funcpara}{N}{int}
    Number of particles
    \end{funcpara}
    \begin{funcpara}{initPosRange}{$(3,2)$ float}
    The initial position regime is given by a min and max for x,y,z
    \end{funcpara}
    \begin{funcpara}{initMomRange}{$(3,2)$ float}
    The initial relativistic momentum regime is given by a min and max for x,y,z
    \end{funcpara}
    \begin{funcpara}{initEmissionTimeRange}{$(1,2)$ float}
    The initial emission time regime is given by a min and max value greater than zero.
    \end{funcpara}
    \textit{EX}: \texttt{ps = genRandInitPhasespace(bI.N, [-1,1;-1,1;-1,1], [-1,1;-1,1;-1,1]*1e-3, [0,2])}
    \end{function}
\end{parameter}

\begin{parameter}{field}{struct}
    A struct containing the information for a static or dynamic, electric or magnetic field. In addition to the mesh and field data, some additional information is precomputed and saved as part of each field struct to speed up the simulation. Therefore, it is not recommended to manipulate field parameters arbitrarily, since part of the data is matched.

    \vspace{0.5em}
    \begin{parameter}{field.data.x}{$(nx,ny,nz)$ float}
    Field x-data as a three-dimensional matrix for each mesh node.
    \end{parameter}
    \begin{parameter}{field.data.y}{$(nx,ny,nz)$ float}
    Field y-data as a three-dimensional matrix for each mesh node.
    \end{parameter}
    \begin{parameter}{field.data.z}{$(nx,ny,nz)$ float}
    Field z-data as a three-dimensional matrix for each mesh node.
    \end{parameter}
    \begin{parameter}{field.frq}{float}
    The frequency of the field, if it is $Dynamic$, otherwise it has no effect.
    \end{parameter}
    \begin{parameter}{field.ignoreBoxList}{list\{$(3,1)$ float\}}
    List of boxes in which particles do not interact with the field.
    \end{parameter}
    \begin{parameter}{field.meanCellSize}{$(3,1)$ float}
    The mean mesh spacing in x,y, and z directions.
    \end{parameter}
    \begin{parameter}{field.mesh.x}{$(n,1)$ float}
    Vector of the hex mesh positions in x the field data is defined on. Already contains the \texttt{trans} offset.
    \end{parameter}
    \begin{parameter}{field.mesh.y}{$(n,1)$ float}
    Vector of the hex mesh positions in y the field data is defined on. Already contains the \texttt{trans} offset.
    \end{parameter}
    \begin{parameter}{field.mesh.z}{$(n,1)$ float}
    Vector of the hex mesh positions in z the field data is defined on. Already contains the \texttt{trans} offset.
    \end{parameter}
    \begin{parameter}{field.meshBounds}{$(3,2)$ float}
    Min/max size of the field in x,y,z.
    \end{parameter}
    \begin{parameter}{field.meshIndexLUT.x}{$(N,1)$ int}
    Lookup table to indices for the acceleration structure in x.
    \end{parameter}
    \begin{parameter}{field.meshIndexLUT.y}{$(N,1)$ int}
    Lookup table to indices for the acceleration structure in y.
    \end{parameter}
    \begin{parameter}{field.meshIndexLUT.z}{$(N,1)$ int}
    Lookup table to indices for the acceleration structure in z.
    \end{parameter}
    \begin{parameter}{field.meshIndexLUT.xSpacing}{float}
    Equal distant spacing of the acceleration structure in x. 
    \end{parameter}
    \begin{parameter}{field.meshIndexLUT.ySpacing}{float}
    Equal distant spacing of the acceleration structure in y. 
    \end{parameter}
    \begin{parameter}{field.meshIndexLUT.zSpacing}{float}
    Equal distant spacing of the acceleration structure in z. 
    \end{parameter}
    \begin{parameter}{field.meshIsEquiDist}{$(3,1)$ boolean}
    True if the mesh has equal distant spacing in either x,y, or z direction. 
    \end{parameter}
    \begin{parameter}{field.meshX}{$(nx,ny,nz)$ float}
    Matrix representation of the mesh along x as generated by the Matlab function \texttt{ndgrid()}. Already contains the \texttt{trans} offset. 
    \end{parameter}
    \begin{parameter}{field.meshY}{$(nx,ny,nz)$ float}
    Matrix representation of the mesh along y as generated by the Matlab function \texttt{ndgrid()}. Already contains the \texttt{trans} offset. 
    \end{parameter}
    \begin{parameter}{field.meshZ}{$(nx,ny,nz)$ float}
    Matrix representation of the mesh along z as generated by the Matlab function \texttt{ndgrid()}. Already contains the \texttt{trans} offset. 
    \end{parameter}
    \begin{parameter}{field.name}{String/ char vector}
    Arbitrary user-defined name. 
    \end{parameter}
    \begin{parameter}{field.nx}{int}
    The number of mesh nodes in the x-direction. When \texttt{type} is $Dynamic$ the \texttt{phaseshift} is included. 
    \end{parameter}
    \begin{parameter}{field.ny}{int}
    The number of mesh nodes in the y-direction. When \texttt{type} is $Dynamic$ the \texttt{phaseshift} is included. 
    \end{parameter}
    \begin{parameter}{field.nz}{int}
    The number of mesh nodes in the z-direction. When \texttt{type} is $Dynamic$ the \texttt{phaseshift} is included. 
    \end{parameter}
    \begin{parameter}{field.phaseshift}{float}
    A global phaseshift of the field data. When \texttt{type} is $Dynamic$ it is directly applied to the field data, otherwise, it has no effect.
    \end{parameter}
    \begin{parameter}{field.trans}{$(3,1)$ float}
    Offset vector of the field from the origin in x,y,z direction. Already contained/precomputed into all \texttt{mesh} variables.
    \end{parameter}
    \begin{parameter}{field.type}{[Static, Dynamic]+[E,B,H] char vector}
    Type of the field. 
    \end{parameter}

    \vspace{0.2em}
    \begin{function}{field = importHDF5\_CST(filePath,fieldType,fieldName,trans,scalersField,frq,phaseShift,unitScalFac)}
    Import a field from an HDF5 file and set all additional settings. Currently, it is assumed that the HDF5 file is generated by CST (tested with versions 2022 and 2023). 
    
    \vspace{0.2em}
    \begin{funcpara}{fieldPath}{String/ char vector}
    Path to the HDF5 file, including the full file name.
    \end{funcpara}
    \begin{funcpara}{fieldType}{char vector}
    Set the field type, this has to match the file, otherwise, an error is thrown.
    \end{funcpara}
    \begin{funcpara}{fieldName}{String/ char vector}
    Arbitrary user-defined name. 
    \end{funcpara}
    \begin{funcpara}{trans}{$(3,1)$ float}
    Offset vector of the field from the origin in x,y,z direction.
    \end{funcpara}
    \begin{funcpara}{scalersField}{$(3,1)$ float}
    Scaling of the field strength in x,y,z direction. 
    \end{funcpara}
    \begin{funcpara}{frq}{float}
    Frequency of the field, if it is $Dynamic$, otherwise it has no effect.
    \end{funcpara}
    \begin{funcpara}{phaseShift}{float}
    a phase shift of the field, if it is $Dynamic$, otherwise it has no effect.
    \end{funcpara}
    \begin{funcpara}{unitScalFac}{float}
    Overwrite the mesh scaling unit factor (Exp: mm $\to$ $10^{-3}$) from the ones used by CST.
    \end{funcpara}
    \textit{EX}: \texttt{fieldE = importHDF5\_CST('fieldE.h5','StaticE','MyField',[0,0,0],[10,10,5])} \\
    \textit{EX}: \texttt{fieldDE = importHDF5\_CST('fieldDE.h5','DynamicE','DF0',[-1,1,0],[1,1,1],1e3,0,1e-3)}
    \end{function}
\end{parameter}

\begin{parameter}{fieldList}{list}
    A list of all electric and magnetic fields used in the simulation.
    
    \vspace{0.5em}
    \begin{function}{fieldList = \{fieldE, fieldDE\}}
    \end{function}
\end{parameter}

\begin{parameter}{genSettings}{struct}
    A struct containing all general and miscellanies settings of the solver and simulation. 

    \vspace{0.5em}
    \begin{parameter}{genSettings.axisNumberSlowKill}{int}
    A number that determines which axis (x,y,z) are used for calculating the slow velocity cut-off. Specified as an integer, where 1 corresponds to the x-axis, 2 to the y-axis, and 3 to the z-axis. To use multiple axes, enter the number in a two or three-digit number. Example: '2' only the y-axis, '13' only x- and z-axis, and '123' all axes.
    \end{parameter}
    \begin{parameter}{genSettings.doInteraction}{boolean}
    Set to true to enable particle-particle interaction between particles in the same bunch.
    \end{parameter}
    \begin{parameter}{genSettings.dt\_update\_res}{int}
    Number of simulation steps between updating the time step $dt$. Numbers smaller than one are set to one.
    \end{parameter}
    \begin{parameter}{genSettings.interactionEnergieCutoff\_eV}{float}
    Set the mean kin. energy limit of a particle bunch in eV. This determines when the two-particle interaction is no longer calculated during the simulation. This saves computing time for high-energy particles since the particle interaction only plays a subordinate role. Default: inf.
    \end{parameter}
    \begin{parameter}{genSettings.interactionType}{["Class", "BunchRel"] char vector}
    Select the type of interaction between particles in the same bunch. Class: Assume a purely static electrical interaction; BunchRel: Lorentz transformed field interaction from a quasi-static electric interaction in a stretched local frame according to the mean particle velocity. See Chap. \ref{chp:Docu_eeInterac}
    \end{parameter}
    \begin{parameter}{genSettings.killBoxList}{list of $(3,2)$ float}
    List of boxes that kill particles inside. Each box is represented as a $(3,2)$ matrix for a min, max in x,y,z direction. \\
        \textit{EX}: \texttt{genSettings.killBoxList = \{[-1,1; -0.1,0.1; 0.5,0.7]\}}
    \end{parameter}
    \begin{parameter}{genSettings.killPartInKillBox}{boolean}
    If true particles get killed, when they are inside a killBox. If all particles are dead the simulation will end early.
    \end{parameter}
    \begin{parameter}{genSettings.killPartOutsideFields}{boolean}
    If true particles get killed, when they are outside all fields. If all particles are dead the simulation will end early.
    \end{parameter}
    \begin{parameter}{genSettings.killSlowParts}{boolean}
    If true, particles slower than the mean of the K fastest particles times the cut-off factor\\ (\texttt{.slowVelCutOffFactor}) get killed. 
    \end{parameter}
    \begin{parameter}{genSettings.numberOfFastPart}{int}
    The number of fastest particles used for calculating the velocity to calculate the cut-off velocity.
    \end{parameter}
    \begin{parameter}{genSettings.saveEverNStep}{int}
    The amount of steps between data is saved during the simulation to \texttt{simState}, where \texttt{1} represents each step.
    \end{parameter}
    \begin{parameter}{genSettings.saveFieldInWindow}{boolean}
    Save each time step save the field inside the window into \texttt{simState}. Caution!
    \end{parameter}
    \begin{parameter}{genSettings.saveFields}{boolean}
    At the beginning, save all fields once to the \texttt{simState}.
    \end{parameter}
    \begin{parameter}{genSettings.showInfo}{boolean}
    If true some additional information gets printed to the console. The amount of additional Information may vary from version to version.
    \end{parameter}
    \begin{parameter}{genSettings.slowVelCutOffFactor}{float}
    A number between 0 and 1 defines the slow velocity cut-off.
    \end{parameter}
    \begin{parameter}{genSettings.solver}{[General, General\_CInterpo, HPC] char vector}
    Set the solver used for simulation, see Chap. \ref{chp:Docu_Solvers} for more information on each option.
    \end{parameter}
    \begin{parameter}{genSettings.visBoxList}{list of $(3,2)$ float}
    List of boxes for visual reference in the trajectory plots. Each box is represented as a $(3,2)$ matrix for a min, max in x,y,z direction. This does not affect the simulation. \\
        \textit{EX}: \texttt{genSettings.visBoxList = \{[-1,1; -0.1,0.1; 0.5,0.7]\}}
    \end{parameter}
\end{parameter}

\begin{parameter}{timeInfo}{struct}
    A struct containing all time-related parameters for the simulation. 

    \vspace{0.5em}
    \begin{parameter}{timeInfo.Nt}{int}
    The number of simulation steps. 
    \end{parameter}
    \begin{parameter}{timeInfo.dt}{int}
    Fallback size of time step in seconds. Intended use varies with the function \textit{dtFunc} used.
    \end{parameter}
    \begin{parameter}{timeInfo.dtFunc}{Function handle}
    A function that will be evaluated during the simulation to determine the next time step length $dt$. The function is called with the following input parameters: \texttt{phasespace}, \texttt{bunchInfo}, \texttt{fieldList}, \texttt{windowInfo}, \texttt{stepCounter}, \texttt{currentTime}, \texttt{olddt}, \texttt{dtVarargin} and has to return one float: \texttt{dt}. Alternatively, the function can also be formulated using \textit{varargin}.
    
    \vspace{0.2em}
    \begin{function}{dt = equalTimeSteps(varargin\{phasespace, bunchInfo, fieldList, windowInfo, \\stepCounter, currentTime, olddt, dtVarargin\})}
    A function that returns the first element of \texttt{dtVarargin} as new \texttt{dt}, thus realizing a constant time step function. During the simulation the function is called with the following list of input parameters in this order: phasespace, bunchInfo, fieldList, windowInfo, stepCounter, currentTime, olddt, dtVarargin
    
    \vspace{0.2em}
    \begin{funcpara}{phasespace}{struct}
    Current phase space.
    \end{funcpara}
    \begin{funcpara}{bunchInfo}{struct}
    Current bunchInfo.
    \end{funcpara}
    \begin{funcpara}{fieldList}{list of field structs}
    Current field list.
    \end{funcpara}
    \begin{funcpara}{windowInfo}{struct}
    Current window info.
    \end{funcpara}
    \begin{funcpara}{stepCounter}{float}
    The number of the current simulation step.
    \end{funcpara}
    \begin{funcpara}{currentTime}{float}
    The current simulated time.
    \end{funcpara}
    \begin{funcpara}{olddt}{float}
    Previous time step length.
    \end{funcpara}
    \begin{funcpara}{dtVarargin}{list}
    Additional time step function parameters. 
    \end{funcpara}
    \end{function}
    \end{parameter}
    \begin{parameter}{timeInfo.dtVarargin}{list}
    A list of dtFunc-function depended parameters.
    \end{parameter}
\end{parameter}

\begin{parameter}{window}{struct}
    Some solvers (exp. \texttt{General}) use a window for each field to minimize the grid data interpolated to each particle position. 

    \vspace{0.5em}
    \begin{parameter}{window.margin}{$(3,2)$ int}
    The additional number of margin cells to each window in x,y, and z in the positive and negative direction.  
    \end{parameter}
    \begin{parameter}{window.maxWindowSize}{$(3,1)$ int}
    X,y,z cell number limit for the window. EX: \texttt{window.maxWindowSize = [inf, inf, inf]}
    \end{parameter}
    \begin{parameter}{window.mode}{[Adaptive, PreCalc, None] char vector]}
    Select the possible operation modes when updating the window. \texttt{Adaptive}: Update the window depending on the positions of the particles, such that all particles are inside plus a margin. This mode needs the fields: \texttt{.velMargin}, \texttt{.margin}, \texttt{.maxWindowSize}. \texttt{PreCalc}: The windows are updated depending on a list of pre-calculated windows depending on time stamps. This mode needs the fields: \texttt{.timeStamp}, \texttt{.windows}.  \texttt{None}: The window will not be updated, by default this results in windows with the same size as the field mesh.
    \end{parameter}
    \begin{parameter}{window.timeStamps}{list of floats}
    List of times when to switch to the corresponding window in \texttt{.windows}. Only necessary when using the \textit{PreCalc} mode.
    \end{parameter}
    \begin{parameter}{window.updateRate}{int}
    The number of steps between window updates.
    \end{parameter}
    \begin{parameter}{window.velMargin}{float}
    The mean velocity of the particles times the \texttt{velMargin} clamped to the mesh of the field will be added as a margin to the window in the direction of the mean velocity. This will enlarge the window in the direction of mean particle velocity. This can help increase the number of needed time steps between window updates but proceed with caution.
    \end{parameter}
    \begin{parameter}{window.windows}{list of $(3,2,n)$ int}
    List of pre-calculated windows, where each set of windows is defended as a matrix of x,y,z min, max indices for each field in the same order as in \texttt{fieldList}. Only necessary when using the \textit{PreCalc} mode. 
    \end{parameter}
\end{parameter}

\begin{parameter}{plotOptions}{struct}
    Specify general settings for the existing plot scripts.

    \vspace{0.5em}
    \begin{parameter}{plotOptions.energyIneV}{boolean}
    If True the energy plots will be converted to units of keV instead of J.
    \end{parameter}
    \begin{parameter}{plotOptions.plotOverTime}{boolean}
    If True the graphs will be plotted over time instead of the number of the simulation step.
    \end{parameter}
\end{parameter}

\subsection{Solvers} \label{chp:Docu_Solvers}

Currently, there are three solvers implemented each focusing on a different regime: \texttt{General}, \texttt{General\_CInterpo}, \texttt{HPC}. In the following subsections, each solver is introduced as well as its pros, cons, and limitations.

\begin{figure}[ht!]
	\center	
    \includegraphics[width=0.95\textwidth]{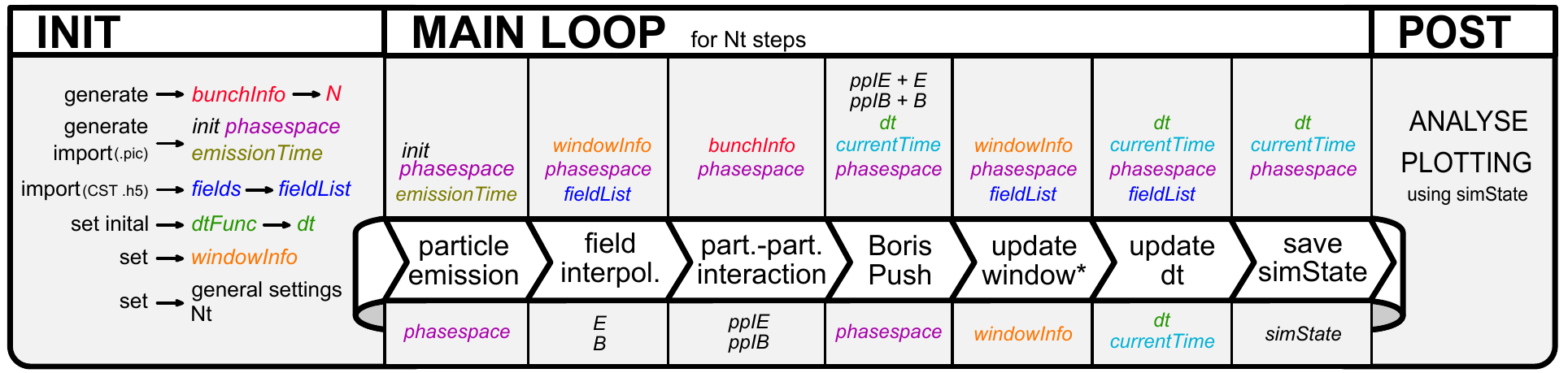}
    \caption{Simplified visualization of the solver initialization, main loop, and post-processing. Each section of the main loop represents a substep of the solver. The input data is listed above and the output data is below. *) The "update window" step is only carried out if required. The "kill Particles" step is missing in this visualization, it takes place after the Boris p.}
	\label{fig:Docu_CodeFlow}
\end{figure}

Each solver implements the following steps, which represent one simulation step (see also fig. \ref{fig:Docu_CodeFlow}):

\begin{enumerate}
  \item add particles to the simulation
  \item interpolate the field to the particle positions
  \item calculate the particle-particle-interaction
  \item do the Boris Kick
  \item kill particles 
  \item update windows
  \item calculate the next time step $dt$
  \item save the current time step to \texttt{simState}
\end{enumerate}

\subsubsection{General} \label{chp:Docu_SolversGeneral}

This is the most general solver and as such supports all features of the femtoTrack framework. It uses the windowing of fields functionality to accelerate the computation of the field interpolation. Therefore this solver adds a window update step after the killing of the particles. 

In addition, this solver gives an online update of the elapsed and approximate remaining time, as well as the number of particles remaining. The remaining time approximation assumes that all remaining time steps are as long as the current one and is therefore not reliable when the particles experience major changes in the number or spread.

Probably the most expensive part of each time step is the field interpolation, which scales poorly with the size of the field or window under consideration but scales well with the number of particles.

\subsubsection{General\_CInterpo} \label{chp:Docu_SolversCGeneral}

This solver is a variation on the \texttt{General} solver, as such it also supports all features. But its significant difference is the algorithm used for the field interpolation. It uses a hard-coded linear interpolation in combination with an acceleration structure for fast particle localization inside of a field. Therefore, the computational effort of this approach is more or less constant with higher resolution grids but scales worse with the number of particles compared to the \texttt{General} solver. 

\subsubsection{HPC} \label{chp:Docu_SolversHPC}

\begin{warning}
    This solver needs the Matlab Parallel Computing Toolbox!
\end{warning}

HPC stands for High-Performance Computing or High Parallelized Computing and tries to get extra performance by splitting the work across multiple pool workers. Each worker gets a separate particle bunch to compute, as bunches cannot interact with each other, allowing for parallel computation. Each worker needs a copy of the full fields, among other things, so more memory is needed.

Since the other solvers already largely use the normal Matlab acceleration for matrix calculations, this solver is not guaranteed to be faster. In addition, this solver is not feature equivalent to the other two, so among other things, the functionality of killing particles during the simulation is missing.

\subsection{SimState} \label{chp:Docu_SimStat}

The result of a simulation is the \texttt{simState}. It is a Matlab vector of a struct containing all information about the simulation from the inputs to the data generated by the solver. 
The structure consists of two parts; the \textit{global} information, which is only defined once at the beginning, and the information in each stored step. \\

To access global information from the \texttt{simState}: \texttt{simState(1).global}\, . \\
To access simulation step data: \texttt{simState($\langle time\,step \rangle$)} \,. \\
The description of all contained fields can be found in Chap. \ref{chp:Docu_SimStatStructure}. \\

Save the data to a file using the \texttt{save} function.

\subsubsection{Internal Structure} \label{chp:Docu_SimStatStructure}

\begin{parameter}{simState(1).global}{struct}
    Contains all globally set parameters such as the settings.

    \vspace{0.5em}
    \begin{parameter}{.bunchInfo}{struct}
    Contains the bunch information. See Chap. \ref{chp:Docu_Settings}.
    \end{parameter}
    \begin{parameter}{.fields}{list of field structs or boolean}
    Contains the list of the full field structs, when the setting \texttt{genSettings.saveFields} is true. Otherwise, it is set to \texttt{false}.
    \end{parameter}
    \begin{parameter}{.genSettings}{struct}
    Contains the general information input struct. See Chap. \ref{chp:Docu_Settings}. 
    \end{parameter}
    \begin{parameter}{.psInfo}{struct}
    Contains additional information about the phase space such for example the particle emission times.
    \end{parameter}
    \begin{parameter}{.timeInfo}{struct}
    Contains the time information input struct. See Chap. \ref{chp:Docu_Settings}.
    \end{parameter}
    \begin{parameter}{.window}{struct}
    Contains the settings struct of the window. See Chap. \ref{chp:Docu_Settings}.
    \end{parameter}
    \begin{parameter}{.version}{$(1,3)$ int}
    A version number of the used solver a the time of the simulation. It is split into three parts which should ruffly represent the main version, the subversion, and the small changes/bug fixes. 
    \end{parameter}
\end{parameter}

\begin{parameter}{simState($\langle time\,step \rangle$)}{struct}
    Contains the simulation result that changes each time step.

    \vspace{0.5em}
    \begin{parameter}{.phasesapce}{struct}
    Contains the full phase space of all particle positions and relativistic momenta; as well as the number of particles. The matrices contained therein for, among other things, the position and momentum always have the size corresponding to the complete number of particles. Entries for particles that have either not yet been emitted or are already dead are filled with \texttt{NaN}. See Chap. \ref{chp:Docu_Settings}.
    \end{parameter}
    \begin{parameter}{.curTime}{float}
    The current time of the simulation step is in seconds.
    \end{parameter}
    \begin{parameter}{.dt}{float}
    The next time step length is in seconds.
    \end{parameter}
    \begin{parameter}{.windowInfo}{$(3,2,nFields)$ int}
    The windows used at the time step in the simulation are given by a x,y,z min, and max indices for each field. 
    \end{parameter}
    \begin{parameter}{.windowedField}{struct}
    When \texttt{genSettings.saveFieldInWindow} is true, save the field structs cut to the current window. Otherwise \texttt{false}. The user is should not be required to use this function for reasons of memory space and computational effort and instead is prompted to determine this information from the fields and the also saved window in a postprocessing step.
    \end{parameter}
    \begin{parameter}{.debugTime}{$(1,8)$ float}
    A vector containing the computation time of different parts of the solver: particle emission, field to particle interpolation, particle-particle interaction, Boris push, killing particles, update window, calculating next dt, and time for saving the simulation step.
    \end{parameter}
\end{parameter}

\subsection{Particle-particle Interaction} \label{chp:Docu_eeInterac}

\subsubsection{Stationary Mode} \label{chp:Docu_eeInteracClass}

In this mode, the particle-particle interaction occurs in the same way as if the particles were entirely stationary. This assumption leads to purely electrostatic forces. As such, it only provides a good approximation for very slowly moving particles. Its advantage is in slightly faster calculation time. This mode is active when \path{genSettings.interactionType = 'Class';}.

\subsubsection{Bunch Lorentz-Transformation Mode} \label{chp:Docu_eeInteracBunch}

In this mode the mean velocity~$\bm{\mean{v}}_i$ of the particle bunch $i$ with $k=1 ... N_i$ particles is used for a Lorentz transformation into the comoving frame. Since the bunches are independent, we will drop the index $i$ in the notation, keeping in mind that $N_i$ is different for each bunch. An electrostatic interaction is calculated in this frame, where the space in $\bm{\mean{v}}$-direction is stretched. When transforming back to the laboratory frame, i.e., calculating the inverse Lorentz transform of the 4D electromagnetic field tensor, an additional magnetic field appears. This approximation is exact as long as the relative motion within a bunch is small.

First, using the particle momentum $p_k$ ($k = 1 ... N$) the mean particle velocity $\bm{\mean{v}}$ and the corresponding Lorentz factor $\gamma$ is calculated 
\begin{align}
    \mean{\bm{p}} &= \frac{\sum_{k=1}^{N} \bm{p_\text{k}}}{N} \,, \\
    \gamma &= \sqrt{1 + \left(\frac{\mean{p}}{ m c_0 }\right)^2} \,, \\
    \mean{\bm{v}} &= \frac{ \mean{\bm{p}}}{m \gamma} \,.
\end{align}
Where $m$ is the particle rest-mass and $c_0$ is the speed of light. For the following calculations the general boost matrix
\begin{align}
    \bm{A}(\bm{v}) &= \begin{bmatrix}
        \gamma & -\gamma v_x / c & -\gamma v_y / c & -\gamma v_z / c \\
        -\gamma v_x / c & 1+\left(\gamma -1\right) v_x^2 / v^2 & \left(\gamma -1\right) v_x v_y / v^2 & \left(\gamma -1\right) v_x v_z / v^2\\
        -\gamma v_y / c & \left(\gamma -1\right) v_y v_x / v^2 & 1+\left(\gamma -1\right) v_y^2 / v^2 & \left(\gamma -1\right) v_y v_z / v^2\\
        -\gamma v_z / c & \left(\gamma -1\right) v_z v_x / v^2 & \left(\gamma -1\right) v_z v_y / v^2 & 1+\left(\gamma -1\right) v_z^2 / v^2\\
    \end{bmatrix} 
\end{align}
is determined. Second, the transformation to the comoving frame (marked with an $'$) is performed by stretching the space between particles ($x_k$ position of particles) along the mean bunch velocity using
\begin{align}
    \bm{x}'_k &= \left( \bm{x}_k - \bm{\mean{x}} \right) \cdot [A_{22}(\mean{\bm{v}}),A_{33}(\mean{\bm{v}}),A_{44}(\mean{\bm{v}})]^T ,
\end{align}
where $\mean{x}=\sum_{k=1}^N \bm{x}_k$.
Then, in this frame, the electrostatic field is calculated by 
\begin{align}
    \bm{E'_{k_1,k_2}} &= \frac{e}{4 \pi \epsilon_0} \frac{\left( \bm{x_}{k_2} - \bm{x}_{k_1} \right)}{\left( x_{k_2} - x_{k_1} \right)^3} \, ,
\end{align}
with the vacuum permittivity $\epsilon_0$. The magnetic field due to residual particle motion is neglected. Third, the electromagnetic field tensor~$\bm{F}^{\mu\nu}$ with zero for the magnetic field 
\begin{align}
    \bm{F^{\mu\nu}} &= \begin{bmatrix}
         0  & E_x/c & E_y/c & E_z/c \\
        E_x/c &  0  & -B_z & B_y \\
        E_y/c & B_z &  0  & -B_x \\
        E_z/c & -B_y & B_x & 0 \\
    \end{bmatrix} \to
    \bm{F'}^{\mu\nu} = \begin{bmatrix}
         0  & E'_x/c & E'_y/c & E'_z/c \\
        E'_x/c &  0  & 0 & 0 \\
        E'_y/c & 0 &  0  & 0 \\
        E'_z/c & 0 & 0 & 0 \\
    \end{bmatrix} 
\end{align}
is transformed back into the lab frame using
\begin{align}
    \bm{F} &= \bm{A}^T(\mean{\bm{v_i}}) {\bm{F}'} \bm{A}(\mean{\bm{v_i}})  \,.
\end{align}
Finally, the electric and magnetic field acting on one particle caused by all other particles is summed and added to the external field.

This approach results in a good approximation of the interaction between charged particles even in the highly relativistic regime. However, if the particles fly in opposite directions, this approach no longer works, since the average particle velocity vector no longer reflects the general behavior of all particles.  This mode is active when \path{genSettings.interactionType = 'BunchRel';}.

\subsection{Inside the FTLib} \label{chp:Docu_InsideFTlib}

\subsubsection{Fast Particle Localization in an inhomogeneous Mesh} \label{chp:Docu_InsideFTlibFast}

\begin{figure}[h]
    \centering
    \includegraphics[width=0.5\textwidth]{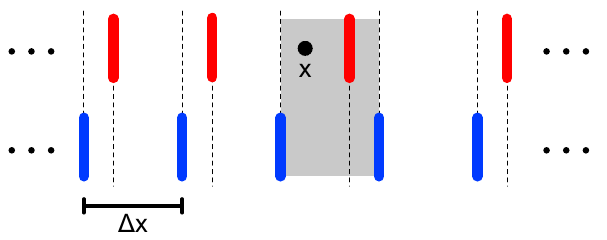}
    \caption{Sketch of the two grids, on the top the inhomogeneous grid on which the field data is defined and on the bottom the additional grid for the faster localization of the particles. }
    \label{fig:Docu_DualGrid}
\end{figure}

The main problem in calculating the field at a particle position is finding all indices to the relevant mesh nodes needed for the field interpolation, especially when working with an inhomogeneous mesh. To address this issue, a simple acceleration structure for each axis of each field is used. Key to this method is a look-up table consisting of $\tilde N_{i} = \frac{\max_{k}(x_{i,k})-\min_{k}(x_{i,k})}{\min_{k}(x_{i,k}-x_{i,k+1}))}$ indices for each grid direction $i \in \{x,y,z\}$, where $x_{i,k}$ are the gird node positions of the original grid with $k = 1 ... N_{i}$. The lookup table is treated as a homogeneously spaced secondary grid which is overlaid with the original grid, see Fig.~\ref{fig:Docu_DualGrid}. Each node contains the corresponding index of the node in the original grid with the next larger position value respectively. This solves the problem of finding the necessary grid nodes surrounding any point inside an inhomogeneously spaced grid by projecting it to a homogeneous grid on which the localization is negligible in computational cost, i.e. it can only be left or right from the respectively localized node on the original grid. 

\subsubsection{Working with Fields} \label{chap:WorkingFields}

This section contains some general and useful information about working with electric or magnetic fields in this solver framework. The implemented Boris method works with electric, vector \textbf{E} and magnetic, vector \textbf{B} fields. The correct import and construction of the used field structs is essential for the correct function of the solver.

\subsubsection{Generation} \label{chap:WorkingFieldsGen}

Generating usable fields is currently only possible via an analytical generation or import from CST as an HDF5 file. It follows a list of the available functions:

\vspace{1em}
\begin{function}{field = genFieldFromFunc(x,y,z,dataFunc,fieldType,trans,scalers, useMeshPos)}
Generate an analytical field using a function handle. 

\vspace{0.2em}
\begin{funcpara}{x}{$(1,n)$ float}
Positions of the mesh nodes in x.
\end{funcpara}
\begin{funcpara}{y}{$(1,n)$ float}
Positions of the mesh nodes in y.
\end{funcpara}
\begin{funcpara}{z}{$(1,n)$ float}
Positions of the mesh nodes in z.
\end{funcpara}
\begin{funcpara}{dataFunc}{function handle}
The generator function, takes three vector float or int inputs for x,y, and z.
\end{funcpara}
\begin{funcpara}{fieldType}{[Static,Dynamic]+[E,B]}
Type of the field.
\end{funcpara}
\begin{funcpara}{trans}{$(3,1)$ float}
Translation of the field in x,y, and z in meters.
\end{funcpara}
\begin{funcpara}{scalers}{$(3,1)$ float}
Scaling of the field data in x,y,z.
\end{funcpara}
\begin{funcpara}{useMeshPos}{boolean}
Sets the input type expected by the \texttt{datafunc} function handle. When false, the number of mesh nodes in x,y, and z is used as input. If true, the position vectors in x,y, and z are used as input.
\end{funcpara}
\textit{EX}: \texttt{field = genFieldFromFunc(-1:0.05:1,-1:0.05:1,-1:0.05:1,@zeros,'StaticE')} - generates a static electric zero field between -1 and 1 in x,y,z with a spacing of 0.05. 
\end{function}

\vspace{1em}
\begin{function}{field = importHDF5\_CST(filePath,fieldType,fieldName,trans,scalersField,frq,phaseShift,unitScalFac)}
Import a field from an HDF5 field generated by CST. 

\vspace{0.2em}
\begin{funcpara}{filePath}{String/ char vector}
Path to the HDF5 file including the file extension (".h5").
\end{funcpara}
\begin{funcpara}{fieldType}{[Static,Dynamic]+[E,D,B,H]}
Type of the field. When importing D or H fields, automatic global conversion to a corresponding E or B field is done using $\epsilon_0$ or $\mu_0$. Any additional global material properties have to be done manually via the \texttt{scalersField} input.
\end{funcpara}
\begin{funcpara}{fieldName}{String/ char vector}
Arbitrary user-defined name.
\end{funcpara}
\begin{funcpara}{trans}{$(3,1)$ float}
Translation of the field in x,y, and z in meters.
\end{funcpara}
\begin{funcpara}{scalersField}{$(3,1)$ float}
Scaling of the field data in x,y,z.
\end{funcpara}
\begin{funcpara}{frq}{float}
The frequency of the dynamic field; has no effect when the field type is static.
\end{funcpara}
\begin{funcpara}{phaseShift}{float}
Applies a global phase shift to a dynamic field. Has no effect when the field type is static.
\end{funcpara}
\begin{funcpara}{unitScalFac}{float}
The importer tries to extract the scaling of the used mesh from the file, if it fails use this to overwrite the scaling factor.
\end{funcpara}
\end{function}

\subsubsection{Field Instancing} \label{chap:FieldInst}

There are some simulation cases in which it is necessary to have multiple copies of the same field at different locations in space. To minimize the memory footprint, it is possible to create multiple instances of a field. This is realized by allowing each field to have multiple translation vectors; one for each instance. The use of instances helps optimize the simulation. This is because each field instance is treated separately when checking if any particles are inside the field boundaries. This is often done when, for example, interpolating the field to the particle positions. Also, when the used solver uses a window, each field instance gets a window.

The system has the following limitations: first, instances of a field represent the same field data and field size. Field ignore boxes are not shared between instances. But this allows only some instances to be affected by an ignore box. 
It has to be taken special care to keep all internal field instance data in sync with each other. This is necessary to ensure correct behavior. Therefore, it is not advised to manually manipulate instanced fields.

\vspace*{2mm}

\begin{function}{field = createInstancedArrayOfField(field, numOfInsts, eachCopyTrans, \\useFieldSizeTrans, instOfInstances)}
Create instanced copies of a field along a given vector. It is possible to instance an already instanced field.

\vspace{0.2em}
\begin{funcpara}{field}{field struct}
The original field struct can already have instances.
\end{funcpara}
\begin{funcpara}{numOfInsts}{int}
Number of copies, including the original. So 1 does not create any instances.
\end{funcpara}
\begin{funcpara}{eachCopyTrans}{$(3,1)$ float}
The translation between successive copies in m (default).
\end{funcpara}
\begin{funcpara}{useFieldSizeTrans}{boolean}
If set to TRUE, the eachCopyTrans input is in units of the field size. Default: False
\end{funcpara}
\begin{funcpara}{instOfInstances}{boolean}
If set to TRUE, also already existing field instances get instanced. Otherwise only the original field is used. Use this to easily generate a 2D array of the same field. Default: false
\end{funcpara}
\textit{EX}: \texttt{field1 = createArrayOfField(field, 3, [0.1,0.2,-0.3])\}} - create two field instances \\
\textit{EX}: \texttt{field2 = createArrayOfField(field1, 2, [1,2,3], true)\}} - the offset of the field instances depends on the field size\\
\textit{EX}: \texttt{field3 = createArrayOfField(createArrayOfField(field, 3, [0,1,0], true), 3, [1,0,0],\\ true, true)\}} - create a perfect 3 by 3 field array in x and y
\end{function}


\subsubsection{Overlapping Field} \label{chap:WorkingFieldsOver}

The default behavior for any two fields of the same kind (electric or magnetic) that overlap in space is to be added together (superposition principle). 

If a second field represents the same field but at a higher resolution, it is possible to generate such a relationship between the two fields. To accomplish this, use the \texttt{addLocalFieldOverwrite} function. Internally, it is handled by adding an ignore box to the coarse/ dominated field of the size of the fine/ dominating field. As the name implies, particles inside this will ignore the coarse/ dominated field and thus interact only with the fine/ dominating field. To make this technique of overlapping fields possible, the user has to guarantee that the field data on the border matches. Currently, there is no way to smoothly interpolate between fields in case of a mismatch. 

Doing this will only guarantee the correct behavior when both fields are added to the simulation. The generated relationship is only internal and currently, the solver will not throw an error if only one of the fields is present in the simulation.

\vspace{1em}
\begin{function}{
lowField = addLocalFieldOverwrite(highField, lowField, replaceList, transferAllBoxes)}
Allows to overlap a locally finer field over another field.

\vspace{0.2em}
\begin{funcpara}{highField}{field struct}
The finer/ dominating field.
\end{funcpara}
\begin{funcpara}{lowField}{field struct}
The coarse/ dominated field.
\end{funcpara}
\begin{funcpara}{replaceList}{boolean}
If set to true, the existing \texttt{ignoreBoxList} of \texttt{lowField} will be replaced by the new one. Otherwise, the new boxes get appended to the list, which is the default behavior. Default: false.
\end{funcpara}
\begin{funcpara}{transferAllBoxes}{boolean}
In addition to the boundary box of the \texttt{highField} all ignore boxes in \texttt{highField.ignoreBoxList} get added to \texttt{lowField.ignoreBoxList}. There may be a case this is useful (I have not found them yet). Default: false.
\end{funcpara}
\textit{EX}: \texttt{lowField = addLocalFieldOverwrite(highField, lowField)} - the highField locally overwrites the lowField, so that the particles interact only with one of the two fields at a time
\end{function}

\subsection{Bunch Generation} \label{chp:BunchGen}

The grouping of particles in bunches, which defines the possibility of interaction between particles, is one of the central concepts of this framework. It allows for the simultaneous simulation of different-sized bunches, for a better statistical representation of, for example, a real-world experiment (Gaus distribution). Each particle therefore can only be in one bunch, this allows all particles in a bunch to also be grouped in the phase space vectors.

There are already three bunch generator functions - for an equal, a Gaussian, and a Poisson distribution of the number of particles in bunches - made available. They can be found in the sub-folder \textit{PS\_Bunch\_gen}. For special particle bunch constellations, the user is advised to combine the creation of the initial phase space and the bunching info struct into the same function. 

\vspace{1em}
\begin{function}{bunchInfo = genEqualBunches(NB, M)}
Generate the information for bunches with the same number of particles.

\vspace{0.2em}
\begin{funcpara}{NB}{Int}
The number of particles per bunch.
\end{funcpara}
\begin{funcpara}{M}{Int}
The number of bunches.
\end{funcpara}
\end{function}

\begin{function}{bunchInfo = genGausBunches(M, meanVal, stdDiv, maxPartPerBunch, fillRest)}
Generate the information for bunches with a \href{https://en.wikipedia.org/wiki/Normal_distribution}{Gaussian/ Normal distribution \ExternalLink} of the number of particles. 

$f(x) = \frac{1}{\sigma * \sqrt{2\pi}} e^{-\frac{1}{2} \frac{x-\lambda}{\sigma}}$

\vspace{0.2em}
\begin{funcpara}{M}{Int}
Maximum number for bunches to generate.
\end{funcpara}
\begin{funcpara}{meanVal}{float}
Mean value of the Gaussian distribution.
\end{funcpara}
\begin{funcpara}{stdDiv}{float}
The standard deviation of the Gaussian distribution.
\end{funcpara}
\begin{funcpara}{maxPartPerBunch}{int}
The maximum number of particles inside a bunch to consider during the creation of bunches.
\end{funcpara}
\begin{funcpara}{fillRest}{boolean}
If false, the number of generated bunches is smaller than or equal to \texttt{M}, such that the particle number distribution of the bunches matches as well as possible to a Gaussian distribution. If true (default), additional bunches, till \texttt{M} bunches are created, are added in such a way that the RMS error is minimal.
\end{funcpara}
\end{function}

\begin{function}{bunchInfo = genPoissonBunches(M, meanVal, maxPartPerBunch, fillRest)}
Generate the information for bunches with a  \href{https://en.wikipedia.org/wiki/Poisson_distribution}{Poisson distribution \ExternalLink} of the number of particles.

$f(k = \{0,1,2,.. \}) = \frac{\lambda^k e^{-\lambda}}{k!}$

\vspace{0.2em}
\begin{funcpara}{M}{Int}
Maximum number for bunches to generate.
\end{funcpara}
\begin{funcpara}{meanVal}{float}
Mean value of the Poisson distribution.
\end{funcpara}
\begin{funcpara}{maxPartPerBunch}{int}
The maximum number of particles inside a bunch to consider during the creation of bunches.
\end{funcpara}
\begin{funcpara}{fillRest}{boolean}
If false, the number of generated bunches is smaller than or equal to \texttt{M}, such that the particle number distribution of the bunches matches as well as possible to a Gaussian distribution. If true (default), additional bunches, till \texttt{M} bunches are created, are added in such a way that the RMS error is minimal.
\end{funcpara}
\end{function}

\subsection{Variable Time Step \texttt{dt}} \label{chp:VarTimestep}

It is possible to have a user-defined variable time step throughout the simulation depending on the current state of the simulation. For this, a function that calculates the next time step length \texttt{dt} must be given. This function handle has to take the following parameters as input and only return a single float. 

\vspace{1em}
\begin{function}{dt = equalTimeSteps(varargin\{phasespace, bunchInfo, fieldList, windowInfo, stepCounter, \\currentTime, olddt, dtVarargin\})}
The function handle given by the user in \texttt{dtFunc} has to work with these input parameters. Alternatively, to shorten the parameter list, it is also possible to use  \texttt{varargin}.
    
\vspace{0.2em}
\begin{funcpara}{phasespace}{struct}
Current phase space.
\end{funcpara}
\begin{funcpara}{bunchInfo}{struct}
Current bunchInfo.
\end{funcpara}
\begin{funcpara}{fieldList}{list of field structs}
Current field list.
\end{funcpara}
\begin{funcpara}{windowInfo}{struct}
Current window info.
\end{funcpara}
\begin{funcpara}{stepCounter}{float}
The number of the current simulation step.
\end{funcpara}
\begin{funcpara}{currentTime}{float}
The current simulated time.
\end{funcpara}
\begin{funcpara}{olddt}{float}
Previous time step length.
\end{funcpara}
\begin{funcpara}{dtVarargin}{list}
Additional time step function parameters. 
\end{funcpara}
\end{function}

\subsubsection{Example: using the CFL-Condition} \label{chp:VarTimestep_CFL}

The Courant–Friedrichs–Lewy (CFL) condition is the number of cells a particle can traverse in one time step. This can be reserved to calculate a corresponding time step $\Delta t$ when given a CFL-number $k_{CFL}$ based on the field mesh using:

\begin{align}
    \Delta t = k_\text{CFL} \frac{\min_\text{i,k} ( \Delta x_\text{i,k} ) }{ v_\text{max}}
\end{align}


Using the smallest cell dimension $\Delta x_{i,k}$ along $i = \{ x,y,z \}$ with $j = 1 ... N_i$ and the fastest particle velocity $v_\text{max}$. Furthermore, only cells within the particle cloud are taken into account when determining the smallest cell dimension.

An implementation of this as code can be found in the file:  \path{/FTLib/dt_fucs/adaptiveTimeSteps_CFL.m}. To use this for the simulation \path{timeInfo.dtFunc = @adaptiveTimeSteps_CFL} has to be set.
It is optimized to only consider mesh cells inside a field mesh cell aligned bounding box around all particles when calculating the smallest cell dimension.

\subsection{Plotting using \texttt{simState}} \label{chp:plotting}

As an essential part of the post-processing of a simulation, the plotting and visualizing of the generated data is important. In this framework, the complete set of data generated by the simulation is saved in the \texttt{simState} struct. For plotting it can be easier to first extract the data of interest and, if necessary, calculate any additional information before plotting them. In the following code segment this is demonstrated; be advised, that it can be necessary to permutate the extracted data for plotting. 

\begin{lstlisting}[language=Matlab, style=myCustomMatlabStyle]
  time=[simState.curTime];
  phaseSpace=[simState.phaseSpace];
  
  for jj = 1:N
    plot(10^12*time, phaseSpace(jj).pos(:,1), '-');
    plot(10^12*time, phaseSpace(jj).pos(:,2), '-');
    plot(10^12*time, phaseSpace(jj).pos(:,3), '-');
  end
\end{lstlisting}

\end{document}